\documentclass[parskip, 12pt]{article}

\usepackage{graphics}
\usepackage{amssymb}
\usepackage{amsmath, framed}
\usepackage[toc,page,header]{appendix}

\usepackage{wrapfig}
\usepackage{lscape}
\usepackage[paper=portrait,pagesize]{typearea}
\usepackage{ltablex}
\usepackage{tabularx}

\usepackage{blkarray}

\usepackage{braket}
\usepackage{epsfig}
\usepackage{amsmath}
\usepackage{amsfonts}
\usepackage{amssymb}
\usepackage{enumerate}
\usepackage{ifpdf}
\usepackage{	qcircuit}
\usepackage{pifont}
\usepackage{graphicx}
\usepackage{braket}
\usepackage{pifont}
\usepackage{array}
\usepackage{longtable}
\usepackage{bm}
\usepackage{hyperref}
\usepackage{color}
\usepackage{makeidx}
\usepackage{mathtools}
\usepackage{bbm}
\usepackage{wrapfig}
\usepackage[skip=0pt]{caption}
\usepackage{titlesec}
\usepackage{amsmath}
\usepackage{tikz}
\usepackage{mathdots}
\usepackage{yhmath}
\usepackage{cancel}
\usepackage{color}
\usepackage{siunitx}
\usepackage{array}
\usepackage{multirow}
\usepackage{amssymb}
\usepackage{gensymb}
\usepackage{booktabs}
\usetikzlibrary{fadings}
\usepackage{hyperref}
\usetikzlibrary{cd}
\usepackage{amsthm}

\usepackage{hhline}

\usepackage{color,soul}

\theoremstyle{remark}

\theoremstyle{definition}

\usepackage[a4paper, margin=2cm, marginparwidth=0cm, marginparsep=0cm, outer=0cm, headheight=11pt, marginpar=0pt]{geometry}

\usepackage{xcolor}
\definecolor{GreenDark}{RGB}{000,179,000}

\usepackage{tikz}
\usepackage{tikz-3dplot}
\usepackage{pgfplots}
\pgfplotsset{compat = 1.3}
\usetikzlibrary{arrows}
\usetikzlibrary{calc}

\def\centerarc[#1](#2)(#3:#4:#5){%
	\draw [#1] ($(#2)+({#5*cos(#3)}, {#5*sin(#3)})$) arc (#3:#4:#5)%
}

\tdplotsetmaincoords{70}{150}

\pgfmathsetmacro{\zOneRot}{10}
\pgfmathsetmacro{\xRot}{15}
\pgfmathsetmacro{\zTwoRot}{30}

\pgfmathsetmacro{\xRot}{10}
\pgfmathsetmacro{\yRot}{15}
\pgfmathsetmacro{\zRot}{30}

\usepackage[font=footnotesize]{caption}

\newcommand{\R}{\mathbb{R}}
\newcommand{\C}{\mathbb{C}}

\newcommand{\ketpsi}{\ket{\psi}}
\newcommand{\kz}{\ket{0}}
\newcommand{\ko}{\ket{1}}
\newcommand{\bz}{\bra{0}}
\newcommand{\bo}{\bra{1}}

\newcommand{\bellzz}{\frac{\ket{00} + \ket{11}}{\sqrt{2}}}

\newcommand{\trace}{\text{Tr}}

\usepackage[utf8]{inputenc}
\usepackage[english]{babel}
 
\usepackage[
backend=biber,
style=numeric,
sorting=ynt
]{biblatex}
 

\titleformat{\chapter}[display]
    {\normalfont\LARGE\bfseries}{\chaptertitlename\ \thechapter}{5pt}{\Large}
\titlespacing*{\chapter}{0pt}{-20pt}{5pt}
\titlespacing*{\section}{0pt}{8pt}{8pt}
\titlespacing*{\subsection}{0pt}{8pt}{8pt}

\DeclarePairedDelimiterX{\inp}[2]{\langle}{\rangle}{#1, #2}
\begin{document}

%\nocite{*}
\newpage
%\title{abc}
\title{QDataset: Quantum Datasets for Machine Learning}
\author{Elija Perrier, Akram Youssry, Chris Ferrie}
\maketitle

\begin{abstract}
    The availability of large-scale datasets on which to train, benchmark and test algorithms has been central to the rapid development of machine learning as a discipline and its maturity as a research discipline. Despite considerable advancements in recent years, the field of quantum machine learning (QML) has thus far lacked a set of comprehensive large-scale datasets upon which to benchmark the development of algorithms for use in applied and theoretical quantum settings. In this paper, we introduce such a dataset, the QDataSet, a quantum dataset designed specifically to facilitate the training and development of QML algorithms. The QDataSet comprises 52 high-quality publicly available datasets derived from simulations of one- and two-qubit systems evolving in the presence and/or absence of noise. The datasets are structured to provide a wealth of information to enable machine learning practitioners to use the QDataSet to solve problems in applied quantum computation, such as quantum control, quantum spectroscopy and tomography. Accompanying the datasets on the associated GitHub repository are a set of workbooks demonstrating the use of the QDataSet in a range of optimisation contexts. 
\end{abstract}

\tableofcontents

\pagenumbering{arabic}

\hypersetup{
    colorlinks,
    citecolor=black,
    filecolor=black,
    linkcolor=black,
    urlcolor=black
}

\newpage

\section{Introduction}
\subsection{Motivating QML Datasets}
 Quantum machine learning (\textbf{QML}) is an emergent multi-disciplinary field combining techniques from quantum information processing, machine learning and optimisation to solve problems relevant to quantum computation \cite{schuld_evaluating_2018, verdon_universal_2018, biamonte_quantum_2017, Amin_Andriyash_Rolfe_Kulchytskyy_Melko_2018}. The last decade in particular has seen an acceleration and diversification of QML across a rich variety of domains. As a discipline at the interface of classical and quantum computing, subdisciplines of QML can usefully be characterised as to where they lie on the classical-quantum spectrum of computation \cite{aimeur_machine_2006}, ranging from quantum-native (using only quantum information processing) and classical (using only classical information processing) to hybrid quantum-classical (a combination of both quantum and classical). At the conceptual core of QML is the nature of how quantum or hybrid classical-quantum systems can \textit{learn} in order to solve or improve results in constrained optimisation problems. The type of machine learning of relevance to QML algorithms very much depends on the specific architectures adopted. Thus QML combines concepts and techniques from quantum computation and classical machine learning, while also exploring novel \textit{quantum} learning architectures.
 
 While quantum-native QML is a burgeoning and important field, the more commonplace synthesis of machine learning concepts with quantum systems arises in classical-quantum hybrid architectures \cite{vidick_quantum_2016, schuld_evaluating_2018, verdon_quantum_2019, zhou_machine_2017}. Such architectures are typically characterised by a classical parametrisation of quantum systems or degrees of freedom (measurement distributions or expectation values) which are updated according to classical optimisation routine. In applied laboratory and experimental settings, hybrid quantum-classical architectures remain the norm primarily due the fact that most quantum systems rely upon classical controls \cite{dong_quantum_2010, viola_dynamical_1999}. To this end, hybrid classical-quantum QML architectures which are able to optimise classical controls or inputs for quantum systems have wider, more near-term applicability for both experiments and NISQ \cite{preskill_quantum_2018, bharti_noisy_2021} devices. Recent literature on hybrid classical-quantum algorithms for quantum control \cite{youssry_beyond_2020, perrier_quantum_2020}, noisy control \cite{youssry_modeling_2020} and noise characterisation \cite{youssry_beyond_2020} present examples of this approach. Other recent approaches include the hybrid use of quantum algorithms and classical objective functions for natural language processing \cite{lorenz_qnlp_2021}. Thus the search for optimising classical-quantum QML architectures are well-motivated from a theoretical and applied perspective.
 
 Despite the increasing maturity of hybrid classical-quantum QML as a discipline, the field lacks many of the characteristics that have been core to the extraordinary successes of classical machine learning in general and deep learning in particular. Classical machine learning has been driven to a considerable extent by the availability of large-scale, high-quality accessible datasets against which algorithms can be developed and tested for accuracy, reliability and scalability. The availability of such datasets as MNIST \cite{lecun_mnist_1998}, ImageNet \cite{russakovsky_imagenet_2015}, Netflix \cite{zhou_large-scale_2008} and other large scale corpora has acted as a catalyst to not just innovations within the machine learning community, but also for the development of benchmarks and protocols that have helped guide the field. Such datasets have also fostered important cross-collaborations among disciplines in ways that have advanced classical machine learning. By contrast, QML as a discipline lacks a similarly standardised set of canonical large-scale datasets against which machine learning researchers (along the quantum-classical spectrum) may benchmark their algorithms and upon which to base innovation. Moreover, the absence of such large-scale standardised datasets arguably holds back important opportunities for cross-collaboration among quantum physicists, computer science and other fields.
 
 In this paper, we seek to address this gap in QML research by presenting a comprehensive QML dataset as a dedicated resource designed for researchers across classical and quantum computation to develop and train hybrid classical-quantum algorithms for use in theoretical and applied settings. We name this dataset \textit{QDataSet} and our project the \textit{QML Dataset Project}. The motivation behind the QML Dataset Project is to map out a similar data architecture for the training and development of QML as exists for classical machine learning. 
 
 \subsection{Results and contributions}
 The contributions of our paper are as follows:
 \begin{enumerate}
     \item Presentation of QDataSet for quantum machine learning, comprising multiple rich large-scale datasets for use in training classical machine learning algorithms for a variety of quantum information processing tasks including quantum control, quantum tomography, quantum noise spectroscopy and quantum characterisation;
     \item Presentation of desiderata of QML datasets in order to facilitate their use by theoretical and, in particular, applied researchers; and
     \item Demonstration of using the QDataSet for benchmarking classical and hybrid classical-quantum algorithms for quantum control.
 \end{enumerate}
 
  \subsection{Structure}

The structure of our paper is as follows. \textbf{Part I} of our paper sets out the key objectives of typical problems in quantum machine learning. Its focus is on applied optimisation and problem solving for experimentalists and theoreticians working on applied quantum information processing problems, such as circuit synthesis, quantum control and tomography. It surveys existing literature and resources relating to the availability, architecture and use cases of quantum datasets for machine learning. We begin with a short overview of the role of large-scale classical datasets in the rapid development o machine learning.  \textbf{Part II} sets out key desiderata of quantum datasets for their use in theoretical and applied machine learning. We set out a taxonomy of features and architectural characteristics to guide the creation, deployment and use of quantum datasets for machine learning. We survey examples across a range of cross-disciplinary contexts, including quantum chemistry, quantum tomography and quantum circuits, together with an analysis of typical quantum datasets used in leading quantum software and programming platforms, such as Qutip, Quantum Tensorflow, Strawberry Fields, Qiskit, Q\# and others \cite{zhao_quantum_2020}. Our objective is to set out a set of standard properties and principles for quantum datasets to enable stakeholders in both quantum computing and classical machine learning, from algorithm designers to experimentalists, to collaborate and translate their research across disciplines more easily.  \textbf{Part III} of the paper is where we introduce and set out the specific taxonomy of the QDataSet. In conjunction with Appendix (\ref{apnd:detail}), we provide a detailed explanation of the data generation, data storage and data architecture for the QDataSet. Because this work is aimed at facilitating cross-disciplinary collaboration, we include in the Appendix an introduction to quantum computing and machine learning concepts for the benefit of researchers for whom such disciplines are beyond their domain expertise. \textbf{Part IV} provides a description of how the QDataSet may be used for hybrid quantum-classical optimisation problems. We focus on identifying how the QDataSet may be used across the development and production pipeline, illustrating the use of the data in preprocessing, in-processing and post-processing contexts. We provide examples of how the QDataSet can be utilised in concert with a suite of classical and hybrid machine learning algorithms to benchmark their performance, including different input and output data, labels and metrics which may be applied. In doing so, our aim is to show machine learning practitioners, in particular, how their domain expertise in solving classical constrained optimisation problems may be applied to the quantum context. \textbf{Part V} concludes our paper and sets out prospective uses of the QDataSet for various research programmes and pathways for the development of QML using large-scale datasets. Before we commence with Part I, we set out a short summary of the QDataSet below. 
 
 \subsection{QDataSet Summary}
 The QDataSet comprises 52 datasets based on simulations of one- and two-qubit systems evolving in the presence and/or absence of noise subject to a variety of controls. It has been developed to provide a large-scale set of datasets for the training, benchmarking and competitive development of classical and quantum algorithms for common tasks in quantum sciences, including quantum control, quantum tomography and noise spectroscopy. It has been generated using customised code drawing upon base-level Python packages in order to facilitate interoperability and portability across common machine learning and quantum programming platforms. Each dataset consists of 10,000 samples which in turn comprise a range of data relevant to the training of machine learning algorithms for solving optimisation problems. The data includes a range of information (stored in list, matrix or tensor format) regarding quantum systems and their evolution, such as: quantum state vectors, drift and control Hamiltonians and unitaries, Pauli measurement distributions, time series data, pulse sequence data for square and Gaussian pulses and noise and distortion data. The total compressed size of the QDataSet (using Pickle and zip formats) is around 14TB (uncompressed, well-over 100TB). Researchers can use the QDataSet in a variety of ways to design algorithms for solving problems in quantum control, quantum tomography and quantum circuit synthesis, together with algorithms focused on classifying or simulating such data. We also provide working examples of how to use the QDataSet in practice and its use in benchmarking certain algorithms. Each part below provides in-depth detail on the QDataSet for researchers who may be unfamiliar with quantum computing, together with specifications for domain experts within quantum engineering, quantum computation and quantum machine learning.
 
As discussed above, the aim of generating the datasets is threefold: (a) simulating typical quantum engineering systems, dynamics and controls used in laboratories; (b) using such datasets as a basis to train machine learning algorithms to solve certain problems or achieve certain objectives, such as attainment of a quantum state $\rho$, quantum circuit $U$ or quantum control problem generally (among others); and (c) enable optimisation of algorithms and spur development of optimised algorithms for solving problems in quantum information, analogously with the role of large datasets in the classical setting. We explain these imperatives below.
\begin{enumerate}
    \item \textit{Datasets as simulations.} Firstly, we have aimed to generate datasets which abstractly simulate the types of data, characteristics and features which would be common across a variety of laboratories and experimental setups. That is, we consider these datasets as abstractions (say of particular Hamiltonians, or noise profiles) which can have any number of physical \textit{realisations}, depending on the experimental design. So different experiments can ultimately realise, in the abstract the same or a sufficiently similar structure as that provided by the data. For example, the implementation of the particular Hamiltonians or state preparation may be done using trapped-ion setups, quantum dot or transmon-based qubits \cite{dewes_characterization_2012}, doped systems or otherwise. We assume the availability of a mapping between the dataset features, such as the controls pulses, and particular control devices (such as voltage or microwave-based controls), for example, in the laboratory. 
    \item \textit{Training algorithms using datasets.} This second aim is related but distinct from the first. The aim is that training models using the datasets has applicability to experimental setups. Thus, for example, a machine learning model trained using the datasets in theory should provide, for example, the optimal set of pulses or interventions needed to solve (and, indeed, optimise) for some objective. It is intended that the output of the machine learning model is an abstraction which can then be realised via the specific experimental setup. The aim then is that the abstraction of each experiments setup allows the application of a variety of machine learning models for optimising in a way that is directly applicable to experimental setups, rather than relying upon experimentalists to then work-out how to translate the model's output into their particular experimental context. Requiring conformity of outputs within these abstract criteria thus facilitates a greater, practical, synthesis between machine learning and the implementation of solutions and procedures in experiments.
    \item \textit{Benchmarking, development and testing.} The third aim of the datasets is to provide a basis for benchmarking, development and testing of existing and new algorithms in quantum machine learning. As discussed above, classical machine learning has historically been characterised by the availability of large-scale datasets with which to train and develop algorithms. The role of these large datasets is multifaceted: (i) they provide a means of \textit{benchmarking} algorithms (see above), such that a common set of problem parameters, constraints and objectives allows comparison among different models; (ii) their size often means they provide a richer source of overt and latent (or constructible) features which machine learning models may draw upon, improving the versatility and diversity of models which may be usefully trained. 
\end{enumerate}

\newpage
\section*{Part I: Objectives and Datasets}
\section{QML Objectives \& Aims}
\subsection{QML Objectives}
Cross-disciplinary programmes focused on building quantum datasets for machine learning will benefit from a framework to categorise and classify the particular objectives of QML architectures and articulation of number of design principles relevant to the taxonomy of QML datasets. 
Designing large-scale datasets for QML requires an understanding of the objectives for which QML research is undertaken and the extent to which those objectives involve classical and/or quantum information processing. 
 Following \cite{aimeur_machine_2006}, the application of machine learning techniques to quantum information processing can be usefully parsed into a simple input / output and process taxonomy on the basis of whether information and computational processes are classical or quantum in nature. Here a process, input or output being `quantum in nature' refers to the phenomenon by which the input or output data was generated, or by which the computational process occurs, is itself quantum in nature given that measurement outcomes are represented as classical datasets from which the existence of quantum states or processes is inferred. Quantum data encoded in logical qubits, for example in quantum states (superpositions or entangled), is different from classical data, in practical terms information about such quantum data arises by inference on measurement statistics whose outcomes are classical. 
\begin{center}
\begin{tabular}{ |p{4cm}||p{4cm}|p{4cm}|p{4cm}|  }
 \hline
 \multicolumn{4}{|c|}{QML Taxonomy} \\
 \hline
 \textbf{QML Division}  & \textbf{Inputs} & \textbf{Outputs} & \textbf{Process} \\
 \hline
 Classical ML   & Classical    & Classical &   Classical \\
 \hline
 Applied classical ML &   Quantum (Classical)  & Classical (Quantum)   & Classical\\
 \hline
 Quantum algorithms for classical problems &Classical & Classical&  Quantum\\
 \hline
 Quantum algorithms for quantum problems    &Quantum & Quantum&  Quantum\\
 \hline
\end{tabular}
\end{center}
% \vspace{12}
This taxonomy can be usefully partitioned into four quadrants depending on the objectives of the QML task (to solve classical or quantum problems) and the techniques adopted (classical or quantum computational methods).  
\\
\begin{center}

\tikzset{every picture/.style={line width=0.75pt}} %set default line width to 0.75pt        

\begin{tikzpicture}[x=0.75pt,y=0.75pt,yscale=-1,xscale=1]

%Shape: Axis 2D [id:dp9656518550070439] 
\draw  (176.13,265.73) -- (502.13,265.73)(208.73,2.03) -- (208.73,295.03) (495.13,260.73) -- (502.13,265.73) -- (495.13,270.73) (203.73,9.03) -- (208.73,2.03) -- (213.73,9.03)  ;
%Straight Lines [id:da7495201777074723] 
\draw    (351.13,0.03) -- (351.13,265.03) ;

%Straight Lines [id:da5690055076116556] 
\draw    (209.13,137.03) -- (503.13,137.03) ;

% Text Node
\draw (282,290) node  [align=left] {{\fontfamily{ptm}\selectfont Classical }\\{\fontfamily{ptm}\selectfont computation}};
% Text Node
\draw (414,290) node  [align=left] {{\fontfamily{ptm}\selectfont Quantum }\\{\fontfamily{ptm}\selectfont computation}};
% Text Node
\draw (165,190) node  [align=left] {{\fontfamily{ptm}\selectfont Classical }\\{\fontfamily{ptm}\selectfont problems }};
% Text Node
\draw (163,63) node  [align=left] {{\fontfamily{ptm}\selectfont Quantum }\\{\fontfamily{ptm}\selectfont problems }};
% Text Node
\draw (280,190) node  [align=left] {{\fontfamily{ptm}\selectfont Classical}\\{\fontfamily{ptm}\selectfont machine }\\{\fontfamily{ptm}\selectfont learning }};
% Text Node
\draw (435,198) node  [align=left] {{\fontfamily{ptm}\selectfont Quantum}\\{\fontfamily{ptm}\selectfont algorithms}\\{\fontfamily{ptm}\selectfont for classical}\\{\fontfamily{ptm}\selectfont optimisation }};
% Text Node
\draw (281,66) node  [align=left] {{\fontfamily{ptm}\selectfont Applied }\\{\fontfamily{ptm}\selectfont classical }\\{\fontfamily{ptm}\selectfont ML }};
% Text Node
\draw (438,65) node  [align=left] {{\fontfamily{ptm}\selectfont Quantum }\\{\fontfamily{ptm}\selectfont algorithms}\\{\fontfamily{ptm}\selectfont to solve QIP }};

\draw   (351.13, 137.03) circle [x radius= 5, y radius= 5]   ;
\end{tikzpicture}

\end{center}
% \subsubsection{Classical machine learning}
\textit{Classical machine learning for classical data}. The first quadrant (bottom-left) covers the application of classical computational (machine learning) methods to solve classical problems, that is, problems not involving data or processes of a quantum character. 
\textit{Classical machine learning for quantum data}. 
The second (top-left) quadrant covers the application of classical computational and machine learning techniques to solving problems of a quantum character. Specifically, this subdivision of QML covers using  standard machine learning techniques to solving problems specific to the theoretical or applied aspects of quantum computing, including optimal circuit synthesis \cite{youssry_characterization_2020, perrier_quantum_2020, riviello_searching_2015}, design of circuit architectures and so on. Either input or output data are quantum in nature, while the computational process by which optimisation, for example, occurs is itself classical.  \textit{Quantum algorithms for classical optimisation}.
The third quadrant of problem (bottom-right) covers the application of quantum algorithmic techniques to solving classical problems. In this subdivision, algorithms are designed leveraging the unique characteristics of quantum computation, in a way that assist in optimising classical problems or solve certain classes of problems that may be intractable on a classical computer. Quantum algorithms are designed with machine learning characteristics, potentially utilising certain computational resources or processes unavailable when constrained to classical computation. Examples of such algorithms include variational quantum eigensolvers \cite{peruzzo_variational_2014, yang_optimizing_2017, wecker_progress_2015, mcclean_theory_2016}, quantum analogues of classical machine learning techniques (e.g. quantum PAC learning \cite{lloyd_quantum_2014}) and hybrid quantum analogues of deep learning architectures (see \cite{verdon_universal_2018, vidick_quantum_2016, schuld_supervised_2018, An_Zhou_2019} for background). \textit{Quantum algorithms for quantum information processing}.
The fourth (top-right) quadrant covers the application of quantum algorithms to solve quantum problems, that is, problems whose input or output data is itself quantum in nature. This division covers the extensive field of quantum algorithm design, including the famous Grover and Shor algorithms \cite{grover_quantum_1997, brassard_exact_1997, shor_polynomial-time_1997}.  The QDataSet fits within the second subdivision of QML, its primary use being envisaged as in the development of classical algorithms for optimisation problems of engineered quantum systems. Our focus on classical techniques applied to quantum data is deliberate: while advancements in quantum algorithms are both exciting and promising, the unavailability of a scalable fault-tolerant quantum computing system and limitations in hybrid NISQ devices mean that for the vast majority of experimental and laboratory use cases, the application of machine learning is confined to the classical case. Secondly, as a major motivation of this paper is to provide an accessible basis for classical machine learning practitioners to enter the QML field, it makes sense to focus primarily on applying techniques from the classical domain to quantum data.

\subsection{Large-Scale Data and Machine Learning}
Classical machine learning has become one of the most rapidly advancing scientific disciplines globally with immense impact across applied and theoretical domains. The advancement and diversification of machine learning over the last two decades has been facilitated by the availability of large-scale datasets for use in the research and applied sciences. Large-scale datasets \cite{lecun_mnist_1998, deng_imagenet_2009, goodfellow_deep_2016} have emerged in tandem with increasing computational power that has seen the velocity, volume and veracity of data increase \cite{kitchin_what_2016, hastie_elements_2013}. Such datasets have both been a catalyst for machine learning advancements and a consequence or outcome of increasing scope and intensity of data generation. The availability of large-scale datasets led to the evolution of data mining, applied engineering and even theoretical results in high energy physics \cite{albertsson_machine_2018}. 

An important lesson for QML is that developments within these fields have been facilitated using large-scale datasets in a number of ways. Large-scale datasets improve the trainability of machine learning algorithms by enabling finer-grained optimisations via commonplace methods such a backpropagation. This has particularly been true within the field of deep learning and neural networks \cite{goodfellow_deep_2016}, where large-scale datasets have enabled the development of deeper and richer algorithmic architectures able to model complex non-linearities and functional forms, in turn leading to drastic improvements and breakthroughs across domains such as image classification, natural language processing \cite{devlin_bert_2019, brown_language_2020} and time series analysis. With an abundance of data on which to train algorithms, new techniques such as regularisation and dimensionality reduction to address problems arising from large-scale datasets, including overfitting and complexity considerations, have in turn spurred novel innovations that have contributed to the advancement of the field. Large-scale datasets have also provided a \textit{standardising} function by providing a common basis upon which algorithmic performance may be benchmarked and standardised. By providing standard benchmarks, large-scale datasets have enabled researchers to focus on important features of algorithmic architecture in the design of improvements to training regimes. Such datasets have also enabled the fostering of the field via competitive platforms such as Kaggle, where researchers compete to improve upon state of the art results.
\section*{Part II: Dataset Characteristics \& Design}
\section{Quantum Dataset Characteristics}

\subsection{Characteristics of large-scale datasets}
Large-scale classical machine learning datasets share common structural and architectural characteristics designed to facilitate the objectives for which the datasets were compiled. There are a range of considerations including the specific objectives, the types of data to be stored, the degree of structuring of data (including whether highly structured or unstructured), the dimensionality of datasets, the extent of preprocessing of datasets required, data quality issues (such as missing, uncertain or incorrect data - an issue for example in quantum information processing contexts given sources of error and uncertainty), data imputation for missing datasets, visible or hidden data (e.g. whether data is direct or a feature constructed from other data), the number of data points, format, default tasks of the datasets, data temporality (how contemporaneous data is), control of datasets and access to data. Datasets are also structured depending on the machine learning algorithms for which they were developed, taking into account the types of objectives, loss functions, optimisers, development environment and programming languages of interest to researchers. Table (\ref{table:characteristics}) sets out a range of issues and desiderata in this regard.
\begin{center}
\begin{table}
\begin{tabular}{ |p{4cm}||p{12cm}|  }
 \hline
 \multicolumn{2}{|c|}{Data Set Characteristics} \\
 \hline
 \textbf{Item}  & \textbf{Description}  \\
 \hline
 Objectives  & Specification of the objectives for which the dataset was both created and to be used \\
 \hline
 Description  & Sufficient description of data, representation and theoretical description\\
 \hline
  Training/test  & Identification of training (in-sample) and test (out-of-sample) subsets of data\\
 \hline
 Data Types  & Specification of the types of data and formats to be used   \\
 \hline
 Structuring  & Degree to which data is structured or unstructured    \\
  \hline
 Dimensionality  & The dimension of the datasets, dimensional reduction or kernel methods needed   \\
  \hline
 Preprocessing  & Extent to which preprocessing is required in dataset, covering transformations of data such as  sparsification or  decomposition  \\
  \hline
 Data quality, consistency and completeness & Extent to which data is missing, uncertain or incorrect or noisy along with any necessary imputation methods  \\
  \hline
 Visible v. hidden  & Extent to which data points are `direct' or `indirect' (inferred) and whether imputations necessary   \\
 \hline
\end{tabular}
 \caption{Characteristics of large-scale datasets which can guide the generation of QML datasets and development of QML dataset taxonomies.}
 \label{table:characteristics}
\end{table}
\end{center}
Large-scale dataset characteristics affect the utility of the datasets in applied contexts. Such characteristics are relevant to the design of quantum datasets. Below we set out a number of principles used in the design of the QDataSet which we believe provide a useful taxonomy for the QML community to consider when generating data for use in machine learning-based problems. The aim of the proposed taxonomy for quantum datasets is to facilitate their interoperability across machine learning platforms (classical and quantum) and for use in optimisation for experimentalists and engineered quantum systems. While taxonomies and specific architectures will differ across domains, our proposed QDataSet taxonomy we believe will assist the QML and classical ML community to guide large-scale data generation towards principles of interoperability summarised in Table (\ref{table:characteristics}) and explained below:
\begin{enumerate}
    \item \textit{Objectives}. Quantum datasets, as with classical datasets, benefit from being constructed with particular objectives in mind. Most major classical large-scale datasets are compiled for specific objectives such as, for example, classification or regression tasks. In a quantum setting, such objectives include quantum algorithm design, circuit synthesis, quantum control, tomography or measurement-based objectives (such as sampling). The QDataSet's objectives are to provide training data for use in the development of machine algorithms for controlled experimental and engineered quantum systems. This objective has informed the feature selection and structural design, such as inclusion of measurement statistics, Hamiltonian and unitary sequences and the various types of noise and distortion. 
    \item \textit{Description}.  Sufficiently describing the dataset, efficiently representing the data and providing theoretical context for how and why the datasets are so represented enhances the utility of datasets. Representation of data (its form, structure, data types and so on) affect the ease of use and uptake of datasets. For machine learning, optimal data representation is an important aspect of feature learning, representation learning \cite{hamilton_representation_2017}. In this paper, we go to some lengths to describe the various structural aspects of the QDataSet in order to facilitate its uptake by researchers in designing algorithms. We especially have set-out background information for machine learning practitioners who may be unfamiliar with quantum data in an effort to reduce barriers facing cross-disciplinary collaboration. 
    \item \textit{Training and test sets}. Applied datasets for machine learning require training, validation and test sets in order to adequately train algorithms for objectives, such as quantum control. The design of quantum datasets in general, and the QDataSet in particular, has been informed by desirable properties of training sets. These include, for example: (i) \textit{interoperability}, ensuring training set data can be adequately formatted for use in various programming languages (for example storing QDataSet data via matrices, vectors and tensors in Python); (ii) \textit{generalisability}, preprocessing of datasets to improve generalisability of algorithmic results, especially to test or out-of-sample data \cite{hastie_elements_2013} (in the QDataSet, we do this via providing a variety of noise-affected datasets); (iii) \textit{feature smoothing} trained algorithms can often focus on information-rich yet small subspaces of data which, while informative for in-sample prediction, can lead to decreased generalisability across the majority of in- and out-of-sample data lacking such features. Feature smoothing is a technique to coarse-grain features so that less weight is put on rarer though information rich features in order to improve generalisation. In a quantum context, this may involve an iterative process of trial and error that trains datasets and seeks to identify relevant features; alternatively, it may involve using techniques from quantum information theory to classify regions of high and low information content e.g. via entropy measures. 
    \item \textit{Data precision and type}. Data precision and data typing is an important consideration for quantum datasets, primarily to facilitate ease of interoperability between software and applied/hardware in the quantum space. Others considerations can include the degree of precision with which data should be presented. Ideally, quantum data for use in developing algorithms for application in experimental quantum control or measurement scenarios should allow flexibility of data precision to match instruments used in laboratories on a case-by-case basis. For example, the QDataSet precisification of noise degrees of freedom (such as amplitude, mean and standard deviation) have been informed by collaborations with experimental groups. 
    \item \textit{Structuring}. Data structuring, the degree to which data is structured according to taxonomies, is an important characteristic of classical datasets that affects their use and functionality. For quantum datasets, structuring encompasses the types of information that would be included and how that information is categorised. In the selection of real-world applicable datasets, researchers will have a range of choices of salient information to includes, including: theoretical details of the candidate Hamiltonians, details of the physical laboratory setting such as controls, exogenous parameters such as temperature (a significant environmental variable affecting quantum systems), noise or other disturbances; the characteristics of measurement devices and so on. Spectroscopic information, including details of the spectroscopy used, may also be included. What to include and not include will depend upon the particular uses cases and generality (or specificity) of the datasets. In each case, it makes sense for quantum datasets to contain as much useful information as possible such as about parameters, say exogenous environment parameters, or distortion information which may affect measurement devices. Doing so enables algorithms trained on quantum datasets to improve their performance and generalise better. Examples of such information in the QDataSet include details we have included regarding noise profiles and distortion simulations.
    \item \textit{Dimensionality}. The dimensionality of datasets is an important consideration. Large-dimensional datasets often require adaptation in order to facilitate algorithmic learning. This is especially in order to address the ubiquitous curse of dimensionality \cite{bellman_dynamic_1956}, where, as dimensions of datasets increase, algorithms may fail to converge, gradients may vanish or become stuck in barren plateaux. This may occur during the preprocessing stage, in-processing or during post-processing. Techniques such as principal component analysis \cite{hastie_elements_2013}, matrix factorisation \cite{ciliberto_quantum_2018}, feature extraction together with algebraic techniques such as singular value decompositions are all motivated primarily to reduce the dimensionality and complexity of datasets, thereby minimising the hypothesis search space. Moreover, learning algorithms which can efficiently solve problems with sparse datasets often have computational advantages.  Quantum data by its nature rapidly becomes higher-dimensional as the number of qubit or computations resources increases. Such vast search spaces present challenges for QML, such as the barren plateaux problems \cite{mcclean_barren_2018}, the quantum analogue of the vanishing gradient problem in classical machine learning (albeit with differences arising due to exponentially-large search spaces).
    \item \textit{Preprocessing}. Datasets often require or benefit from preprocessing in order to ameliorate problems during training, such as vanishing gradients, bias or problems with convergence. Preprocessing data can include techniques such as sparsification \cite{ravishankar_online_2015} or smoothing or other strategies. For example, for quantum circuit synthesis, ensuring training data samples are drawn from across the Hilbert space of interest rather than limited to subspaces can assist with generalisation (see \cite{perrier_quantum_2020} for a geometric example). In such cases, quantum dataset preparation may benefit from preprocessing to address sparsity concerns (see \cite{marrero_entanglement_2020} for examples and for classical analogues of vanishing gradients \cite{pascanu_difficulty_2013}). 
    \item \textit{Visibility}. Classical machine learning is often concerned with extraction - or development - of features. In many forms of classical machine learning, such as those using kernel methods, or deep learning, features of importance to optimal performance of an algorithm may need to be inferred from the data. Quantum datasets in many ways face such challenges from the outset as quantum data can never be directly observed, rather it must be inferred from measurement statistics. When constructing quantum datasets, the extent to which such inferred (as distinct from directly observed) data will be an important choice. In the QDataSet, we have chosen to include a range of such `hidden' or `inferred' data to assist practitioners with use of the dataset, including the intermediate forms of Hamiltonian, unitaries and other data that is not itself directly accessible but is a by-product of our simulation (accessible via intermediate layers).
\end{enumerate}
Studying features of particular datasets and their use in classical contexts assists in extracting desirable features for large-scale quantum datasets. ImageNet is one of the leading research projects and database architectures for images \cite{deng_imagenet_2009, russakovsky_imagenet_2015, goodfellow_deep_2016}. The dataset is one of the most widely cited and important datasets in the development of machine learning, especially image-classification algorithms using convolutional, hierarchical and other deep-learning based neural networks. The evolution of ImageNet and its use within machine learning disciplines provides a useful guide and comparison for the development of QML datasets in general. ImageNet comprises two main components: (i) a public semi-structured dataset of images together with (ii) an annual competition and workshop. The dataset provides a `ground truth' standardised set of images against which to train categorical image classification algorithms. The competition and workshop provided and continue to provide an important institutional practice driving machine learning development. While beyond the scope of this paper, the development of QML would arguably be considerably assisted by the availability of QML-focused competitions akin to those commonplace within the classical machine learning community. Such competitive frameworks would motivate and drive the development of scalable and generalisable QML algorithms. As is also evident from classical machine learning, competitive formats are also a useful way for laboratories, companies or other projects to leverage the expertise of the diverse machine learning community.

Another example from the machine learning community which can inform the development of QML is Kaggle, a leading online platform for machine learning-based competitions. Kaggle runs competitions where competitors are provided with prediction tasks, training sets and constraints upon the type of algorithm design (such as resource use and so on). Competitors then develop models aiming to optimise a measure of success, such as a standard machine learning metric of accuracy, AUC or some other measure \cite{bojer_kaggle_2021}. The open competitive nature of Kaggle is designed to crowd-source solutions and expertise to problems in machine learning and data science. A `quantum Kaggle' would be of considerable benefit to the QML community by providing a platform through which to spur collaborative and competitive development of quantum algorithms. 

\subsection{QML Datasets}
While quantum datasets for machine learning (quantum and classical) are neither as prevalent nor as ubiquitous as those in the classical realm, there are some examples in the literature of quantum or hybrid quantum-classical datasets generated for use in machine learning contexts. 
QML datasets can be categorised into: (1) \textit{general quantum datasets} produced for purposes other than QML, such as quantum datasets in quantum chemistry of other fields, which can be preprocessed or used as training data in QML contexts. Such datasets are not specifically produced for the purposes of QML per se; (2) dedicated \textit{QML-specific quantum datasets}, generated and structured for the purposes of QML. This second category mainly consists of quantum datasets for use in classical or hybrid machine learning contexts. Quantum datasets currently available tend towards one or other of these classifications, though there is overlap, for example, with quantum datasets designed for use in machine learning which are nevertheless highly domain-specific. Examples include quantum chemistry datasets for use in deep tensor neural networks \cite{schutt_quantum-chemical_2017}, datasets for learning spectral properties of molecular systems \cite{chmiela_machine_2017, ramakrishnan_electronic_2015, rupp_fast_2012, bartok_machine-learning_2013} and for solid-state physics \cite{sutton_crowd-sourcing_2019, nyshadham_machine-learned_2019, szlachta_accuracy_2014,faber_machine_2016}.

A recent example is provided by the dedicated quantum chemistry datasets known as the \textit{QM7-X dataset} \cite{hoja_qm7-x_2021}, an expansive dataset of physiochemical properties for several million equilibrium and non-equilibrium structures of small organic molecules. The QM7-X dataset spans regions of chemical compound space and was generated to provide a basis for machine-learning assisted design of molecules with specific properties. The dataset builds upon previous iterations of QM-series and related quantum chemistry datasets \cite{rupp_fast_2012, ruddigkeit_enumeration_2012, smith_ani-1ccx_2020}. Structurally, the dataset combines global (molecular) properties and local (atomic) information, including ground state quantities (spectra and moments) along with response quantities (related to polarisation and dispersion). The dataset is highly domain-specific and represents a salient example of a dataset designed to spur machine-learning driven research within a particular field. 
\subsection{QML and QC platforms}
The use of quantum datasets in machine learning has been facilitated over the last several years by a surge in quantum programming and languages and platforms for both QML and quantum computing generally. While such platforms are dynamic and changing, it is important that quantum datasets for machine learning be constructed to be as interoperable with platforms in the quantum and classical machine learning community. Generators of large-scale quantum datasets should be cognisant of how their data can be (more easily) used in such platforms below and also how their datasets can be designed in ways that facilitates their ease of use within common machine learning languages, such as TensorFlow, PyTorch and across languages, such as Python, C\#, Julia and others.  The QDataSet has been specifically designed in relatively elementary Python packages such as Numpy in order to facilitate its use across the machine learning community, but also in a way that we hope makes it useable and clearly understandable by the quantum engineering community. We deliberately selected Python as the language of choice within which to build the QDataSet simulation given its status as the leading classical programming language of choice for machine learning. It also is a language adopted across many of the quantum platforms above. We built the QDataSet using Numpy to produce datasets as transferable as possible (rather than, for example, in Qutip). A familiarity with the emerging quantum programming and QML ecosystem is useful for the design of quantum datasets. We set out a few examples of leading quantum programming platforms below.   

\textit{Qutip} \cite{johansson_qutip_2013} is a leading quantum simulation and algorithmic package in Python used for open quantum systems' simulation. The package, while not developed specifically for QML, is widely used for hybrid quantum-classical systems' research. Inputs to Qutip algorithms are Numpy-based vectors, tensors and matrices used to represent density matrices, quantum states and operators. Qutip permits a wide range of simulations to be run and data to be generated, including for state preparation, control and drift Hamiltonians, pulse sequences and noise modelling. As discussed below, the QDataSet, which was build in Python using primarily the Numpy package, but was verified using Qutip. \textit{Q\#} is Microsoft's primary open-source programming language for quantum algorithm design and execution. The platform comprises a number of libraries, simulators and a software development kit (QDK). \textit{Quantum Tensorflow} (QTF) \cite{broughton_tensorflow_2020} is a hybrid quantum-classical version of Google's leading open-source machine learning Tensorflow platform. QTF is constructed to enable the synthesis of classical and quantum algorithmic machine learning, for example classically parameterised quantum circuits, variational quantum circuits and eigensolvers, quantum convolutional neural networks and other quantum analogues of classical machine learning architectures. QTF follows Tensorflow's overall machine learning structure and data taxonomy. Input data is usually in the form of tensors. QTF's in-platform datasets vary depending on use case, but the platform primarily draws upon classical datasets for hybrid use cases (quantum computation applied to solving classical optimisation tasks).  For simulated quantum-native data, QTF draws upon \textit{Cirq}, Google's open source framework for programming quantum computers \cite{cirq_developers_cirq_2021}. Cirq is focused on providing a software library for research into and simulations of quantum circuits, the idea being to develop quantum algorithms in Cirq that can be run on quantum computers and simulators. 

\textit{Strawberry Fields} \cite{killoran_strawberry_2019} is an open-source QML and quantum algorithm programming platform developed by Xanadu for photonic quantum computing. \textit{Qiskit} is another open source software development kit for quantum circuits, control and applications on quantum and hybrid computers  \cite{aleksandrowicz_qiskit_2019}. Qiskit is based on an open source quantum assembly language (QASM) standardised abstraction of quantum circuits. Other platforms enabling the integration of quantum datasets and QML algorithms (either quantum or classical) include those available via IBM's Quantum Experience. The QDataSet has been designed to for interoperability across most of these platforms. Practically speaking, this means that researchers can select dataset features of interest, such as tensors of quantum states, Hamiltonians, unitary operators (gates) or even noise information and integrate as datasets for use in algorithms designed using platforms above. Similarly, machine learning researchers should find the form of data relatively familiar to typical datasets in machine learning, where information is encoded in tensors, lists or matrices. Examples of using similar datasets in customised TensorFlow machine learning models can be found in various sources \cite{youssry_characterization_2020, perrier_quantum_2020}.
\subsection{Quantum data in datasets}
An important aspect of quantum dataset design is the decision regarding what quantum information to include in the dataset. In this section, we list types of quantum data which may be included in large-scale quantum datasets. By \textit{quantum data}, we refer to data generated or characterising quantum systems or processes. Quantum data may comprise a range of different properties, features or characteristics of quantum systems, the environment around quantum systems. It may comprise data and information abstracted into a particular representation or form, such as circuit gates, algebraic formulations, codes etc or more physical forms, such as statistics or read-outs from measurement devices. For QML datasets, it is useful to ensure that quantum data is sufficient for classical machine learning researchers to understand and for integrating quantum data into their algorithmic techniques. For example, a classically parameterised quantum circuit, as common throughout the QML literature, would typically include data or tensors of the relevant parameters, the operators related to such parameters (such as generators) and the quantum states (vectors or density operators) upon which the circuit acts.

\begin{center}
\begin{table}
\begin{tabular}{ |p{4cm}||p{12cm}|  }
 \hline
 \multicolumn{2}{|c|}{\textbf{Quantum Data}} \\
 \hhline{|==|}
 \textbf{Item}  & \textbf{Description}   \\
 \hhline{|=||=|}
  \hline
 Quantum states  & Description of states in computational basis, usually represented as vector or matrix (for $\rho$). May include initial and evolved (intermediate or final) states \\
 \hline
 Measurement operators & Measurement operators used to generate measurements, description of POVM. \\
 \hline
 Measurement distribution  & Distribution of measurement outcome of measurement operators, either the individual measurement outcomes or some average (the QDataSet is an average over noise realisations). \\
 \hline
 Hamiltonians & Description of Hamlitonians, which may include system, drift, environment etc Hamiltonians. Hamiltonians should also include relevant control functions (if applicable). \\
 \hline
 Gates and operators  & Descriptions of gate sequences (circuits) in terms of unitaries (or other operators). The representation of circuits will vary depending on the datasets and use case, but ideally quantum circuits should be represented in a way easily translatable across common quantum programming languages and integrable into common machine learning platforms (e.g. TensorFlow, PyTorch).\\
 \hline
 Noise  & Description of noise, either via measurement statistics, known features of noise, device specifications. \\
 \hline
 Controls  & Specification and description of the controls available to act on the quantum system.\\
 \hline
\end{tabular}
\caption{An example of the types of quantum data features which may be included in a dedicated large-scale dataset for QML. The choice of such features will depend on the particular objectives in question. We include a range of quantum data in the QDataSet, including information about quantum states, measurement operators and measurement statistics, Hamiltonians and their corresponding gates, details of environmental noise and controls.}
\label{table:quantumdatatable}
\end{table}
\end{center}

%%%%%%%%%%%%%%%%%%%%%%%%%%%%%%%%%%%%%%%%%%%%%%%%%%%%%%%%%%%%%%%%%%%%%%%%%%%%%%
%% added by Akram
\section*{QDataSet Taxonomy \& Background}
\section{Quantum information processing}
\subsection{Overview}
In Appendix (\ref{apnd:detail}), we provide an overview of the key elements of quantum information processing of relevance to the use of quantum datasets. We aim to equip classical machine learning practitioners with a minimum working knowledge of the ways in which quantum data and quantum computing differ from their classical counterparts relevant to solving problems in QML. Our focus, as throughout this paper, is on the application of classical and hybrid classical-quantum machine learning algorithms and techniques to solve constrained optimisation problems using quantum data (as distinct from the application of purely quantum algorithms). Our synopsis of quantum postulates and quantum information processing below aims also to provide a mapping between ways in which data and computation is characterised in quantum contexts and their analogues in classical machine learning. For example, the description of dataset characteristics, dimensionality, input features, training data to what constitutes labels, the types of loss functions applicable are all important considerations in classical contexts. By providing a straightforward way to translate quantum dataset characteristics into typical classical taxonomies used in machine learning, we aim to help lower barriers to more machine learning practitioners becoming involved in the field of QML. A high-level summary of some of the types of quantum features that QML-dedicated datasets ideally would contain is set out in Table (\ref{table:quantumdatatable}). Our explication of the QDataSet below provides considerable detail on each quantum data feature contained within the datasets including parameters, assumptions behind our choice of quantum system specifications, measurement protocols and noise context. A dictionary of specifications for each example in the QDataSet is set out in the Appendix in Table (\ref{table:datasetproperties}).   

\subsection{Quantum postulates}
Quantum information processing is characterised by constraints upon how information is represented and processed arising from the foundational postulates of quantum mechanics. In Appendix (\ref{apnd:detail}), we provide an overview of key elements of quantum postulates \cite{nielsen_quantum_2010} and quantum information processing for classical machine learning practitioners solving optimisation problems for quantum engineering. In particular, we describe ways in which quantum data are typically represented (such as via tensors), how quantum processes are usually expressed and technicalities of how hybrid classical-quantum systems in common programming languages. As described below, the QDataSet is constructed to mimic conditions in laboratories and experiments where inputs and outputs to quantum systems are classical, such as via classically characterised controls (pulses, voltages) and measurement outcomes in the form of a classical probability distribution over observable outcomes of measurement. Actual quantum states, coherences and other characteristically quantum features of the system, while considered ontologically extant, are in effect reconstructions conditioned upon classical input and output data. In a machine learning context, this means that the encoding of quantum states, quantum processes (such as unitary evolution) represents the encoding of constraints upon how computation may evolve, rather than the input of actual quantum data. To this end, we follow the data generation protocols set out in \cite{youssry_beyond_2020} which we explicate below.

\subsection{Generated Datasets and Naming convention}
Each dataset can be categorised according to the number of qubits in the system and the noise profile to which the system was subject. Table (\ref{tab:cats}) sets out a summary of such categories. For category 1 of the datasets, we created datasets with noise profiles N1, N2, N3, N4, together with the noiseless case. This gives a total of 5 datasets. For category 2,  the noise profiles for the X and Z respectively are chosen to be (N1,N5), (N1,N6), (N3,N6). Together with the noiseless case, this gives a total of 4 datasets. For category 3 (two-qubit system), we chose only the 1Z (identity on the first qubit, noise along the $z-$axis for the second) and Z1 (noise along the $z-$axis for the first qubit, identity along the second) noise to follow the (N1,N6) profile. This category simulates two individual qubit with correlated noise sources. For category 4, we generate the noiseless, (N1,N5), and (N1,N6) for the 1Z and Z1 noise. This gives 3 datasets. Therefore, the total number off datasets at this point is 13. Now, if we include the two types of control waveforms, this gives a total of 26. If we also include the cases of distortion and non-distorted control, then this gives a total of 52 datasets. Comprehensive detail on the noise profiles used to generate the datasets is contained in Appendix (\ref{apnd:detail}).

We chose a convention for the naming of the dataset to try delivering as much information as possible about the chosen parameters for this particular dataset. The name is partitioned into 6 parts, separated by an underscore sign ``\_''. The first part is either the letter ``G'' or ``S'' to denote whether the control waveform is Gaussian or square. The second part is either "1q" or ``2q'' to denote the dimensionality of the system. The third part denotes the control Hamiltonian. It is formed by listing down the Pauli operators we are using for the control for each qubit, and we separate between qubit by a hyphen ``-''. For example, category 1 datasets will have ``X'', while category 4 with have ``IX-XI-XX''. The fourth part is optional and it encodes the noise Hamiltonian following the same convention of the third part. The fifth which is also optional part contains the noise profiles following the same order of operators in the fourth part. If the dataset is for noiseless simulation, the the fourth and fifth parts are not included. Finally, the sixth part denotes the presence of control distortions by the letter ``D'', otherwise it is empty. For example, the dataset ``G\_2q\_IX-XI-XX\_IZ-ZI\_N1-N6'' is two qubit, Gaussian pulses with no distortions, local X control on each qubit and an interacting XX control along with local Z-noise on each qubit with profile N1 and N6. Another example the dataset ``S\_1q\_XY\_D", is a single-qubit system with square distorted control pulses along X and Y axis, and there is no noise. 

\subsection{Dataset Description and Format}
 
% \subsection{QDataSet Generation}
\subsubsection{QDataSet generation}
 Each dataset in the QDataSet consists of 10,000 examples. An example corresponds to a given control pulse sequence, associated with a set of noise realizations. Every dataset is stored as a compressed zip file, consisting of a number of Python \textit{Pickle} files that stores the information. Each file is essentially a dictionary consisting of the elements described in Table \ref{table:datasetproperties}. We expand upon each element below, building upon and expanding upon the discussion in \cite{youssry_characterization_2020}. The datasets were generated on the University of Technology (Sydney) high-performance computing cluster (iHPC) . The QDataSet was generated on using the iHPC Mars node (one of 30). The node consists of Intel Xeon Gold 6238R 2.2GHz 28cores (26 cores enabled) 38.5MB L3 Cache (Max Turbo Freq. 4.0GHz, Min 3.0GHz) 360GB RAM. We utilised GPU resources using a NVIDIA Quadro RTX 6000 Passive (3072 Cores, 384 Tensor Cores, 16GB Memory). It took around three months to generate over 2020-2021, coming to around 14TB of compressed quantum data. Single-qubit examples were relatively quick (between a few days and a week or so). The two-qubit examples took much longer, often several weeks. 

\subsubsection{QDataSet online}
The QDataSet is available via the Github repository for the datasets \cite{perrier_qdataset_2021} (containing a link to the online Cloudstor database) . The datasets are stored in an online repository and are accessible via links on the site. The largest of the datasets is over 500GB (compressed), the smallest being around 1.4GB (compressed). The QDataSet is provided subject to open-access MIT/CC licensing for researchers globally.

The information is stored following the Tensorflow convention of interpreting multidimensional arrays. For example the noise Hamiltonian for one example is stored as a $(1, M , K , 2 , 2)$ array, where the first dimension is the batch, the second is time assuming $M$ steps, then whatever comes next is related to the object itself. In this case the third dimension denotes the noise realization assuming a maximum of $K$ realizations, and the last two dimensions ensure we have a square matrix of size 2. 

The simulation of the datasets is based on a Monte Carlo method, where a number of evolution trajectories are simulated and then averaged to calculate the observables. The exact details can be found in \cite{youssry_beyond_2020} which we reproduce and expand upon in this paper for completeness. We avoid using master equations to avoid imposing approximations or assumptions that may not be valid or justified. 

\subsubsection{QDataSet Parameters}
Further detail the 52 datasets that we present in this paper for use solving engineering applications discussed in section \ref{sec:applications} using classical machine learning can be found on the repository for the QDataSet \cite{perrier_qdataset_2021}. Table (\ref{table:datasetcharacteristics}) in the Appendix sets out the taxonomy of each example for each of the 52 different datasets (for which there are 10,000 examples each). Each dataset comprises 10,000 examples that are compressed into a Pickle file which is in turn compressed into a zip file. The \textit{Item} field indicates the dictionary key and the \textit{Description} field indicates the dictionary value.

% We fix the evolution time interval to $[0,T]$, the time is discretized into $M$ steps, and we generate $K$ realizations of the noise to average over for the Monte Carlo simulation.

\section*{Part IV}
\section{Optimisation problems for QDataSet}
\subsection{Quantum engineering problems as ML problems}\label{sec:applications}
There are many problems related to the characterization and control of quantum systems that can be solved using ML techniques. In this section, try to give an overview on a number of such problems and how to approach them using ML. Here we provide a brief overview of the different types of problems in quantum computing, engineered quantum systems and quantum control for which the QDataSet and algorithms trained using it may be useful. The list is necessarily non-exhaustive and is intended to provide some direction mainly to machine learning researchers unfamiliar with key problems in applied quantum science.

\subsection{Benchmarking}
 Benchmarking algorithms using standardised datasets is an important developmental characteristics of classical machine learning. Benchmarks provide standardised datasets, preprocessing protocols, metrics, architectural features (such as optimisers, loss functions and regularisation techniques) which ultimately enable research communities to precisify their research contributions and improve upon state of the art results.  Results in classical machine learning are typically presented by comparison with known benchmarks in the field and adjudged by the extent to which they outperform the current state of the art benchmarks. Results are presented in tabular format with standardised metrics for comparison, such as accuracy, F1-score or AUC/ROCR statistics. The role of benchmarking is important in classical contexts. Firstly, it enables a basis for researchers across machine learning subdisciplines to gauge the extent to which their results correlate to algorithmic design as distinct from unique features of training data or use cases. Secondly, it provides a basis for better assessing the algorithmic state of the art within subfields.
 Given its relative nascency, QML literature tends to focus on providing proof-of-concept examples as to how classical, hybrid or quantum-native algorithms can be used for classification or regression tasks. There is little in the way of systematic benchmarking of QML algorithms against their classical counterparts in terms of performance of specifically machine learning algorithms.
 
 Recent examples in a QML setting of benchmarking include comparisons of using different error distributions relevant to quantum chemistry (and how these affect performance) \cite{pernot_impact_2020}, benchmarking machine learning algorithms for adaptive phase estimation \cite{costa_benchmarking_2021} and generative machine learning with tensor networks \cite{wall_generative_2021}. In quantum information science more broadly, comparison with classical algorithms is often driven from computational complexity considerations and the search for quantum supremacy or outperformance, namely whether there exists a classical algorithm which can achieve results with equivalent efficiency of the quantum algorithm. QML research programmes would benefit from adopting (and adapting) practices common in classical machine learning when reporting results, especially the inclusion of benchmarks against leading state of the art algorithms for particular use-cases, such as classification or regression tasks. Selecting the appropriate benchmarking algorithms itself tends to benefit from domain expertise. The QDataSet has been designed in order to be benchmarked against both classical and quantum algorithms.
 
 \subsubsection{Benchmarking by learning protocol}
 Typically machine learning algorithm classification is based firstly on whether the learning protocols are \textit{supervised}, \textit{unsupervised} or semi-supervised \cite{hastie_elements_2013, goodfellow_deep_2016}. \textit{Supervised learning} uses known input and output (label) data to train algorithms to estimate label data. Algorithmic models are updated according to an optimisation protocol, typically gradient descent, in order to achieve some objective, such as minimisation of a loss function that compares the similarity of estimates to label data. \textit{Unsupervised learning}, by contrast, is a learning protocol where label or classification of data is unknown and must be estimated via grouping or clustering together in order to ascertain identifying features. Common techniques include clustering, dimensionality reduction techniques or graph-based methods. \textit{Semi-supervised} learning is an intermediate algorithmic classification drawing on aspects of both supervised and unsupervised learning protocols. Usually known label data, say where only part of a dataset is labelled or classified, is included in architectures in order to learn the classifications which in turn can be used in a supervised contexts. The QDataSet can be used in a variety of supervised, unsupervised or semi-supervised contexts. For example, training an algorithm for optimal quantum control can be undertaken in a supervised context (using pulse data, measurement statistics or Hamiltonian sequences) as label data and modelling estimates accordingly. Alternatively, semi-supervised or unsupervised protocols for tomographic classification can be trained using the QDataSet. In any case, an understanding of standard and state of the art algorithms in each category can provide QML researchers using the QDataSet with a basis for benchmarking their own algorithms and inform the design of especially hybrid approaches (see \cite{schuld_supervised_2018} for an overview and for quantum examples of the above).
 
\subsubsection{Benchmarking by objectives and architecture}
The choice of benchmarking algorithms will also be informed by the objectives and architecture. Classically, algorithms can be parsed into various categories. Typically they are either \textit{regression}-based algorithms, used where the objective is to estimate (and minimise error in relation to) continuous data or \textit{classification}-based algorithms, where the objective is to classify data into discrete categories. \textit{Regression algorithms} are algorithms that seek to model relationships between input variables and outputs iteratively by updating models (and estimates) in order to minimise error between estimates and label data. Typical regression algorithms usually fall within broader families of generalised linear models (GLMs) \cite{gelman_data_2007} and include algorithms such as ordinary least squares, linear and logistic regression, logit and probit models, multivariate models and other models depending on link functions of interest. GLMs are also characterised by regularisation techniques that seek to optimise via penalising higher complexity, outlier weights or high variance. GLMs offer more flexibility for use in QML and for using the QDataSet in particular as they are not confined to assuming errors are normally distributed. Other approaches using Bayesian methods, such as naive Bayes, Gaussian Bayes, Bayesian networks and averaged one-dependence estimators provide yet further avenues for benchmarking algorithms trained on the QDataSet for classification or regression tasks. \textit{Classification} models aim to solve decision problems via classification. They typically compare new data to existing datasets using a metric or distance measure. Examples include clustering algorithms such as k-nearest neighbour, support vector machines, learning vector quantisation, decision-trees, locally weighted learning, or graphical models using spatial filtering. Most of the algorithms mentioned thus far fall within traditional machine learning.

Over the last several decades or so, \textit{neural network} architectures have emerged as a driving force of machine learning globally. Quantum analogues and hybrid neural network architecture has itself a relatively long lineage, including quantum analogues of perceptrons, quantum neural networks, quantum hopfield networks (see \cite{schuld_supervised_2018, dunjko_quantum-enhanced_2016}) through to modern deep learning architectures (such as convolutional, recurrent, graphical and hierarchical neural networks and generative models \cite{goodfellow_deep_2016}).  
One feature of algorithmic development that is particularly important is dealing with the curse of dimensionality - and in a quantum context, barren plateaux. Common techniques to address such problems include dimensionality reduction techniques or symmetry-based (for example, tensor network) techniques whose ultimate goal is to reduce datasets down to their most informative structures while maintaining computational feasibility. While the QDataSet only extends to two-qubit simulations,  the size and complexity of the data suggests the utility of dimensionality-reduction techniques for particular problems, such as tomographic state characterisation. To this end, algorithms developed using the QDataSet can benefit from benchmarking and adapting classical dimensionality-reduction techniques, such as principal component analysis, partial regression, singular value decompositions, matrix factorisation and other techniques \cite{hastie_elements_2013}. It is also important to mention that there has been considerable work in QML generally toward the development of quantum and hybrid analogues of such techniques. These too should be considered when seeking benchmarks.

Finally, it is worth mentioning the use (and importance) of ensemble methods in classical machine learning. Ensemble methods tend to combine what are known as `weak learner' algorithms into an ensemble which, in aggregate, outperforms any individual instance of the algorithm. Each weak learner's performance is updated by reference to a subset of the population of weak learners. Popular examples of such algorithms are gradient-boosting algorithms, such as xGboost \cite{chen_xgboost_2016}.

\subsection{Example applications of the QDataSet}
In this section, we outline a number of applications for which the QDataSet can be used. These include training machine learning algorithms for use in quantum state (or process) tomography, quantum noise spectroscopy and quantum control. Our explication is relatively brief, given full worked examples of the use of QDataSet for such problems is beyond the scope of this paper. However, such applications are the subject of concurrent research programmes. The QDataSet repository contains a number of example Jupyter notebooks.
\subsubsection{Quantum state tomography}
Quantum state tomography involves reconstructing reconstructing an estimate $\hat{\rho}$ of the state of a quantum system given a set of measured observables. The quantum state of interest may be in either a mixed or pure state and the task is to uniquely identify the state among a range of potential states. Tomography requires that measurements be \textit{tomographically complete}, which means that the set of measurement operators form a basis for the Hilbert space of interest. Abstractly, the problem involves stipulating a set of operators $\{O_i\}_i$ as input, and the corresponding desired target outputs $\{\braket{O}_i\}_i$. The objective is to find the best model that fits this data. We know that the relation between these is given by $\braket{O_i}= \trace{(\rho O_i)}$ and we can use this fact to find the estimate of the state. Tomography requires repeatedly undertaking different measurements upon quantum states described by identical density matrices which in turn gives rise to a measurement distribution from which probabilities of observables can be inferred. Such inferred probabilities are used to in turn construct a density matrix consistent with observed measurement distributions thus characterising the state. More formally, assuming an informationally complete POVM $\{O_i \}$ spanning the Hilbert-Schmidt space $B(\mathcal{H})$ of operators on $\mathcal{H}$, we can write the probability of an observation $i$ given density matrix $\rho$ using the Hilbert-Schmidt norm above i.e:
\begin{align}
        p(i|\rho) = \braket{O_i} = \trace(\rho O_i )
    \end{align}
Data are gathered from a discrete set of $M$ experiments, where each experiment is a process of initial state preparation, applying an sequence of gates $\{ G_j \}$ and measuring. This experimental process and measurement is repeated $N$ times leading to a frequency count $n_i$ of a particular observable $i$. The probability of that observable is then estimated as:
 \begin{align*}
            p(i|\rho) \approx \frac{n_i}{N} = \hat{p}_i
        \end{align*}
% We then have:
% \begin{align*}
%     \braket{O_i} = \hat{p}_i
% \end{align*}
from which we reconstruct the density matrix $\rho$. Quantum process tomography is a related but distinct type of tomography. In addition, we also have a set of test states $\{ \rho_j \}$ which span $B(\mathcal{H})$. An unknown gate sequence $G_k$ comprising $K$ gates is applied to the states such that:
 \begin{align}
            p(i|G,\rho_j) \approx \frac{n_i}{N} = \hat{p}_{j,i}
        \end{align}
The QDataSet can be used to train algorithms for machine learning algorithms for tomography. Quantum state and process tomography is particularly challenging. One must ensure that the estimate we get is physical, i.e. positive semidefinite with unit trace. Furthermore, the number of measurements $N$ required for sufficient precision to completely characterise $\rho$ scales rapidly. Each of the $K$ gates in a sequence $G_k$ requires $d^2(d-1)$ (where $d = \dim |B(\mathcal{H})|$) experiments (measurements) to sufficiently characterise the quantum process is $Kd^4-(K-2)d^2 - 1$. Beyond a small number of qubits, it becomes computationally infeasible to completely characterise states by direct measurement, thus parametrised or incomplete tomography must be relied upon. Machine learning techniques naturally offer potential to assist with such optimisation problems in tomography, especially neural network approaches where inherent non-linearities may enable sufficient approximations that traditional tomographic techniques may not. Examples of the use of classical machine learning include demonstration of improvements due to neural network-based (non-linear) classifiers over linear classifiers for tomography tasks \cite{gao_experimental_2018} and classical convolutional neural networks to assess whether a set of measurements is informationally complete \cite{teo_benchmarking_2021}.

The objective of an algorithm trained using the QDataSet may be, for example, be to predict (within tolerances determined by the use case) the tomographic description of a final quantum state from a limited set of measurement statistics (to avoid having to undertake $N$ such experiments for large $N$). Each of the one- and two-qubit datasets is informationally complete with respect to the Pauli operators (and identity) i.e. can be decomposed into a one- and two-dimensional Pauli basis. There are a variety of objectives and techniques which may be adopted. Each of the 10,000 examples for each profile constitutes an experiment comprising initial state preparation, state evolution and measurement. One approach using the QDataSet would be to try to produce an estimate $\hat{\rho}(T)$ of the final state $\rho(T)$ (which can be reconstructed by application of the unitaries in the QDataSet to the initial states) using the set of Pauli measurements $\{ E_O \}$. To train an algorithm for tomography without a full set of $N$ measurements being undertaken, on can stipulate the aim of the machine learning algorithm is then to take a subset of those Pauli measurements as input and try to generate a final state $\hat{\rho}(T)$ that as closely approximates the known final state $\rho(T)$ provided by the QDataSet. 

A variety of techniques can be used to draw from the measurement distributions and iteratively update the estimate $\hat{\rho}(T)$, for example gradient-based updating of such estimates \cite{youssry_efficient_2019}. The distance measure could be any number of the quantum metrics described above, including state or operator fidelity, trace distance of quantum relative entropy. Classical loss functions, such as MSE or RMSE can then be used (as is familiar to machine learning practitioners) to construct an appropriate loss function for minimisation. A related, but alternative, approach is to use batch fidelity where the loss function is to minimise the error between a vector of ones and fidelities, the vector being the size of the relevant batch. Similar techniques may also be used to develop tools for use in gate set tomography, where the sequence of gates $G_k$ is given by the sequence of unitaries $U_0$ in the QDataSet. In that case, the objective would be to train algorithms to estimate $G_k$ given the set of measurements, either in the presence of absence of noise. Table (\ref{table:quantumtomography}) sets out a summary.  

% \subsection*{Quantum State Tomography}
%========tomography table
\begin{center}
\begin{table}[!ht]
\begin{tabular}{ |p{3cm}||p{13cm}|  }
 \hline
 \multicolumn{2}{|c|}{QDataSet features for quantum state tomography} \\
 \hhline{|==|}
 \textbf{Item}  & \textbf{Description}   \\
 \hline
 Objective  & Algorithm to learn characterisation of state $\rho$ given measurements $\{ E_O\}$.  \\
 \hline
 Inputs  & Set of Pauli measurements $\{ E_O\}$, one for each of the $M$ experiments (in the QDataSet, this is  \\
 \hline
 Label  & Final state $\rho(T)$\\
 \hline
  Intermediate inputs  & 
  
  \begin{itemize}
      \item Hamiltonians
      \item Unitary operators
      \item Initial states $\rho_0$
  \end{itemize}
  
  \\
%  \hline
 
%  Intermediate weights  & 
  
%   \begin{itemize}
%       \item a[*]
%   \end{itemize}
  
%   \\
 \hline
 
Output  & Estimate of final state $\hat{\rho}(T)$\\
 \hline
 Metric  & 
 \begin{itemize}
     \item State fidelity $F(\rho,\hat{\rho})$
     \item Quantum relative entropy
 \end{itemize}
 
 \\
%   \hline
% Pseudocode  & [*] \\
  
 \hline
\end{tabular}
\caption{QDataSet features for quantum state tomography. The left columns lists typical categories in a machine learning architecture. The right column describes the corresponding feature(s) of the QDataSet that would fall into such categories for the use of the QDataSet in training quantum tomography algorithms. }
\label{table:quantumtomography}
\end{table}
\end{center}

\subsubsection{Quantum noise spectroscopy}
The QDataSet can be used to develop and test machine algorithms to assist with noise spectroscopy. In this problem, we are interested in finding models of the noise affecting a quantum system given experimental measurements. In terms of the $V_O$ operators discussed earlier, we would like to find an estimate of $V_O$ given a set of control pulse sequences, and the corresponding observables. The QDataSet provides a sequence of $V_O$ operators encoding the average effect of noise on measurement operators. This set of data can be used to train algorithms to estimate $V_O$ from noisy quantum data, such as noisy measurements or Hamiltonians that include noise terms. An example approach includes as follows and proceeds from the principle that we have known information about quantum systems that can be input into the algorithmic architecture (initial states, controls, even measurements) and we are trying to estimate unknown quantities (the noise profile). The inputs to the algorithm would include: the initial quantum states, in the QDataSet case the initial states (being eigenstates of the Pauli operators). Intermediate inputs would include the system and noise Hamiltonians $H_0,H_1$ and/or the system and noise unitaries $U_0, U_1$. Alternatively, inputs could also include details of the various noise realisations. The type of inputs will depend on the type of applied use case, such how much information may be known about noise sources. Label data would be the set of measurements $\{ E_O \}$ (expectations of the observables. Given the inputs (control pulses) and outputs, the problem becomes estimating the mapping $\{ V_O \}$, such that inputs are mapped to outputs via equation (\ref{eqn:voeqn}). Note that details bout noise realisations or distributions are never accessible experimentally.

Alternatively, architectures may take known information about the system such as Pauli measurements as inputs. Another approach is to adopt a similar architecture to \cite{youssry_characterization_2020, youssry_efficient_2019} and construct a multi-layered architecture that replicates the simulation, where the $\{ \hat{V}_O \}$ are extracted from intermediate or custom layers in the architecture. Such greybox approaches may combine traditional or deep-learning methods and have the benefit of providing finer-grained control over algorithmic structure by allowing, for example, the encoding of `whitebox' or known processes from quantum physics (thereby eliminating the need for the algorithm to learn these processes). Table (\ref{table:quantumspectroscopy}) sets out one approach that may be adopted.

%=========table for noise spectroscopy
\subsection*{Quantum Noise Spectroscopy}
\begin{center}
\begin{table}[!ht]
\begin{tabular}{ |p{3cm}||p{13cm}|  }
 \hline
 \multicolumn{2}{|c|}{QDataSet features for quantum noise spectroscopy} \\
 \hhline{|==|}
 \textbf{Item}  & \textbf{Description}   \\
 \hline
 Objective  & Algorithm to estimate noise operators $\{ V_O\}$, thereby characterising relevant features of noise affecting quantum system.  \\
 \hline
 Inputs  & Pulse sequence, reconstruted from the \textit{pulse\_parameters} feature in the dataset.  \\
 \hline
 Label  & Set of measurements $\{ E_O\}$\\
 \hline
  Intermediate inputs  & 
  
  \begin{itemize}
      \item Hamiltonians
      \item Unitary operators
      \item Initial states $\rho_0$
  \end{itemize}
  
  \\
%  \hline
%  Intermediate weights  & 
  
%   \begin{itemize}
%       \item a[Weight matrix characterising $\hat{V}_O$]
%   \end{itemize}
  
%   \\
 \hline
Output  & Estimate of measurements $\{ \hat{E}_O\}$\\
 \hline
 Metric  & 
 \begin{itemize}
     \item MSE (between estimates and label data)
     \begin{align}
        MSE(E_O,\hat{E}_O)
    \end{align}
 \end{itemize}
 
 \\
%   \hline
% Pseudocode  & [*] \\
  
 \hline
\end{tabular}
\caption{QDataSet features for quantum noise spectroscopy. The left columns lists typical categories in a machine learning architecture. The right column describes the corresponding feature(s) of the QDataSet that would fall into such categories for the use of the QDataSet in training quantum tomography algorithms. }
\label{table:quantumspectroscopy}
\end{table}
\end{center}

\subsubsection{Quantum control and circuit synthesis}
The QDataSet has been designed in particular to facilitate algorithmic design for quantum control. As described in some detail above, we wish to compare different (hybrid and classical) machine learning algorithms to optimise a typical problem in quantum control, namely describing the optimal sequence of pulses in order to synthesise a target unitary $U_T$ of interest. Here the datasets form the basis of training, validation and test sets used to train and verify each algorithm. The target (label) data for quantum control problems can vary. Typically, the objective of quantum control is to achieve a reachable state $\rho(T)$ via the application of control functions to generators, such as Pauli operators. Achieving the objective means developing an algorithm that outputs a sequence of control functions which in turn describe the sequence of experimental controls $f_\alpha(t)$. A typical machine learning approach to quantum control takes $\rho(T)$ as an input together with intermediate inputs, such as the applicable generators (e.g. Pauli operators encoded in the system Hamiltonian $H_0(t)$ of the QDataset). The algorithm must learn the appropriate time-series distribution of $f_\alpha(t)$ (the set of control pulses included in the QDataSet, their amplitude and sequence in which they should be applied) in order to synthesise the estimate $\hat{\rho}(T)$. Some quantum control problems are agnostic as to the quantum circuit pathway (sequence of unitaries) taken to reach $\hat{\rho}(T)$, though usually the requirement is that the circuit be resource optimal in some sense, such as time-optimal (shortest time) or energy-optimal (least energy). 

One approach is to treat $f_\alpha(t)$ as the label data and $\rho(T)$ as input data to try to learn a mapping between them. A naive blackbox approach is unlikely to efficiently solve this problem as it would require learning from scratch solutions to the Schr{\"o}dinger equation. A more efficient approach may be to encode known information, such as the laws governing Hamiltonian evolution etc into machine learning architecture, such as greybox approaches described above. In this case, the target $f_\alpha(t)$ must be included as a intermediate input into the system Hamiltonians governing the evolution of $\rho(t)$, yet remains the output of interest. In such approaches, the input data would be the intial states of the QDataSet with the label data $\rho(T)$ (and end estimate $\hat{\rho}(T)$) being. Applicable loss functions then seek to minimise the (metric) distance between $\rho(T)$ and $\hat{\rho}(T)$, such as fidelity $F(\rho(T),\hat{\rho}(T))$. To recover the sought after sequence $f_\alpha(t)$, the architecture then requires a way to access the intermediate state of parameters representing $f_\alpha(t)$ within the machine learning architecture.

If path specificity is not important for a use case, then  trained algorithms may synthesise any pathway to achieve $\hat{\rho}(T)$, subject to the optimisation constraints. The trained algorithm need not replicate the pathways taken to reach $\rho(T)$ in the training data. If path specificity is desirable, then the QDataSet intermediate operators $U_0(t)$ and $U_1(t)$ can be used to reconstruct the intermediate states i.e. to recover the time-independent approximation:
\begin{align}
    U(t)^\dagger \rho U(t) \approx (U_n...U_1) \rho (U_1...U_n)
\end{align}
An example of such an approach is in \cite{perrier_quantum_2020} where time-optimal quantum circuit data, representing geodesics on Lie group manifolds, is used to train algorithms for generating time-optimal circuits. Table (\ref{table:quantumcontrol}) sets out schemata for using the QDataSet in a quantum control context.

%=========table for quantum control
% \subsection*{Quantum Control}
\begin{center}
\begin{table}[!ht]
\begin{tabular}{ |p{3cm}||p{13cm}|  }
 \hline
 \multicolumn{2}{|c|}{QDataSet features for quantum control} \\
 \hhline{|==|}
 \textbf{Item}  & \textbf{Description}   \\
  \hline
 Objective  & Algorithm to learn optimal sequence of controls to reach final state $\rho(T)$ or (equivalently) synthesise target unitary $U_T$.  \\
 \hline
 Inputs  &   Hamiltonians containing Pauli generators $H_0(t)$ \\
 \hline
 Label  & Final state $\rho(T)$ and (possibly) intermediate states $\rho(t_j)$ for each time-interval $t_j$.\\
 \hline
  Intermediate fixed inputs  & 
  
  \begin{itemize}
      \item Sequence of unitary operators $U_0(t),U_1(t)$
      \item Initial states $\rho_0$
  \end{itemize}
  
  \\
 \hline
 Intermediate weights  & 
  
  \begin{itemize}
      \item Sequence of pulses $f_\alpha(t)$ including parameters depending on whether square or Gaussian (for example)
  \end{itemize}
  
  \\
 \hline

Output  & Estimate of final state $\hat{\rho}(T)$ and intermediate states $\hat{\rho}(t_j)$\\
 \hline
 Metric  & 
 \begin{itemize}
     \item Average operator fidelity $F(\rho,\hat{\rho})$
 \end{itemize}
 
 \\
%   \hline
% Pseudocode  & [*] \\
  
 \hline
\end{tabular}
\caption{QDataSet features for quantum control. The left columns lists typical categories in a machine learning architecture. The right column describes the corresponding feature(s) of the QDataSet that would fall into such categories for the use of the QDataSet in training quantum control algorithms. The specifications are just one of a set of possible ways of framing quantum control problems using machine learning.}
\label{table:quantumcontrol}
\end{table}
\end{center}

% \newpage
% \KOMAoptions{paper=landscape,pagesize, DIV=current}
% \recalctypearea
% % \section{This is my Landscape Page}
% \begin{center}
% \begin{table}
% \hskip-3.0cm
% \begin{tabular}{ |p{3cm}||p{4cm}|p{5cm}|p{4cm}|p{5cm}|p{2cm}| }
%  \hline
%  \multicolumn{6}{|c|}{QDataSet Characteristics} \\
%  \hhline{|======|}
%  \textbf{Problem}  & \textbf{Inputs}   & \textbf{Intermediate} & \textbf{Output / Label} & \textbf{Metrics} & \textbf{Loss}\\
%  \hline
%  Control & Initial states $\rho_0$, Hamiltonian $U_T$ & Generators (e.g. Pauli operators), Pulses $f(t)$ &  Final state $\rho(T)$ & State fidelity, QRE & MSE\\
%  \hline
%  Circuit synthesis  & Initial states $\rho_0$ & Generators (e.g. Pauli operators), Pulses $f(t)$ & Target unitary estimate $\hat{U}_T$  & Operator fidelity & MSE\\
%  \hline
%  Tomography & Pauli measurements  & [*] & Final state estimate $\rho(T)$ & State fidelity; Hamming distance (?)&\\

%  \hline
% \end{tabular}
% \caption{QDataSet characteristics. The left column identifies each item in the respective QDataSet examples (expressed as keys in the relevant Python dictionary) while the description column describes each item. We elaborate on the description of each item below. }
% \label{table:quantumsummary}
% \end{table}
% \end{center}

% \newpage
% \KOMAoptions{paper=portrait,pagesize}
% \recalctypearea

\newpage
\section*{Part V}
\section{Conclusion and future work}
In this work, we have presented the QDataSet, a large-scale quantum dataset available for the development and benchmarking of quantum machine learning algorithms. The 52 datasets in the QDataSet comprise simulations of one- and two-qubit datasets in a variety of noise-free and noisy contexts together with a number of scenarios for exercising control. Large-scale datasets play an important role in classical machine learning development, often being designed and assembled precisely for the purpose of algorithm innovation. Despite its burgeoning status, QML lacks such datasets designed specifically to facilitate QML algorithm development. The QDataSet has been designed to address this need by providing a resource for cross-collaboration among machine learning practitioners, quantum information researchers and experimentalists working on applied quantum systems. In this paper we have also ventured a number of principles which we hope will assist producing large-scale datasets for QML, including specification of objectives, quantum data features, structuring, preprocessing. We set-out a number of key desiderata for quantum datasets in general. We also have aimed to provide sufficient background context across quantum theory, machine learning and noise spectroscopy for machine learning practitioners to treat the QDataSet as a point of entry into the field of QML. The QDataSet is sufficiently versatile to enable machine learning researchers to deploy their own domain expertise to design algorithms of direct use to experimental laboratories.

The scope for the application of the QDataSet in QML research is considerable. QML is an emerging cross-disciplinary field whose progression will benefit from the establishment of taxonomies and standardised practices to guide algorithm development. In this vein, we sketch below a number of research programmes, building upon principles upon which the QDataSet was designed, in order to foster the [development] of QML datasets and research practices.
\begin{enumerate}
    \item \textit{Algorithm development}. The primary research programme flowing from the QDataSet involves its use in the development of algorithms with direct applicability to quantum experimental and laboratory setups. As discussed above, the QDataSet has been designed to be versatile and of use across a range of use cases, such as quantum control, tomography, noise spectroscopy. In addition, its design enables machine learning practitioners to benchmark their algorithms. Future research involving the QDataSet could cover a systematic benchmarking of common types of classical machine learning algorithms for supervised and unsupervised learning. We also anticipate research programmes expanding upon greybox and hybrid models, using the QDataSet as a way to benchmark state of the art QML models.
    \item \textit{Quantum taxonomies}. While taxonomies within and across disciplines will differ and evolve, there is considerable scope for research programmes examining optimal taxonomic structuring of quantum datasets for QML. In this paper, we have outlined a proposed skeleton taxonomy that datasets for QML may wish to adopt or adapt, covering specification of objectives, ways in which data is described, identification of training (in-sample) and test (out-of-sample) data, data typing, structuring, completeness and visibility (see Table (\ref{table:characteristics}). Further research in these directions could include expanding taxonomic classifications of QML in ways that connect with classical machine learning taxonomies, taking the QDataSet as an example. Doing so would facilitate greater cross-collaboration among computer scientists and quantum researchers by allowing researchers to easily transfer their domain expertise.
    \item \textit{Experimental interoperability}. An important factor in expanding the reach and impact of QML is the extent to which QML algorithms are directly applicable to solving problems in applied engineering settings. Ideally, QML results and architecture should be `platform agnostic' - able to be applied to a wide variety of experimental systems, such as superconductor, transmon, trapped ion, photonic or quantum dot-based setups. Achieving interoperability across dynamically evolving technological landscapes is challenging for any discipline. For QML, the more that simulations within common platforms (such as those mentioned above) can easily integrate into each other and usefully simulate applied experiments, the greater the reach of algorithms trained using them. To the extent that the QDataSet can demonstrably be used across various platforms, algorithm design using it can assist these research imperatives.
\end{enumerate}
We encourage participants in the quantum community to advance the development of dedicated quantum datasets for the benefit of QML and expect such efforts to contribute significantly to the advancement of the field and cross-disciplinary collaboration.

\subsection*{Acknowledgements}
Research and development of the QDataSet was supported by the Centre for Quantum Software and Information at the University of Technology, Sydney. This research was supported in part by the iHPC Facility at the University of Technology, Sydney. We specifically acknowledge the assistance of Simon Cruik and Dr Matt Gaston.

\renewcommand{\thesection}{\Alph{section}}

%===========BIBLIOGRAPHY
\printbibliography

%============Appendix
\newpage
\appendix
\addcontentsline{toc}{section}{Appendix}
% \part{Appendix}
\section{QDataSet Characteristics}

%======Longtable example

\begin{center}
\begin{longtable}{|p{4cm}||p{13cm}|}
\hline
\multicolumn{2}{|c|}{QDataSet Characteristics} \\
 \hhline{|==|}
 \textbf{Item}  & \textbf{Description}   \\
 \hline
 \textit{simulation\_parameters}  & \begin{itemize}
 \item \textit{name}: name of the dataset;
     \item \textit{dim}: the dimension $2^n$ of the Hilbert space for $n$ qubits (dimension 2 for single qubit, 4 for two qubits);
     \item $\Omega$: the spectral energy gap;
     \item \textit{static\_operators}: a list of matrices representing the time-independent parts of the Hamiltonian (i.e. drift components);
     \item \textit{dynamic\_operators}: a list of matrices representing the time-dependent parts of the Hamiltonian (i.e. control components), without the pulses. So, if we have a term $f(t) \sigma_x + g(t) \sigma_y$, this list will be $[\sigma_x, \sigma_y]$;
     \item \textit{noise\_operators}: a list of time-dependent parts of the Hamiltonian that are stochastic (i.e. noise components). so if we have terms like $\beta_1(t) \sigma_z + \beta_2(t) \sigma_y$, the list will be $[\sigma_z, \sigma_y]$;
     \item \textit{measurement\_operators}: Pauli operators (including identity) ($I,\sigma_x,\sigma_y, \sigma_z$)'
     \item \textit{initial\_states}: the six eigenstates of the Pauli operators;
     \item \textit{T}: total time (normalised to unity);
     \item \textit{num\_ex}: number of examples, set to 10,000;
     \item \textit{batch\_size}: size of batch used in data generation (default is 50);
     \item $K$: number of randomised pulse sequences in Monte Carlo simulation of noise (set to $K = 2000$);
     \item \textit{noise\_profile}: N0 to N6 (see above);
     \item \textit{pulse\_shape}: Gaussian or Square;
     \item \textit{num\_pulses}: number of pulses per interval;
     \item \textit{elapsed\_time}: time taken to generate the datasets.
 \end{itemize} \\
 \hline
 \textit{pulse\_parameters}  & The control pulse sequence parameters for the example:
 \begin{itemize}
     \item Square pulses: $A_k$ amplitude at time $t_k$;
     \item Gaussian pulses: $A_k$ (amplitude), $\mu$ (mean) and $\sigma$ (standard deviation).
 \end{itemize}
 \\
 \hline
  \textit{time\_range}  & A sequence of time intervals $\Delta t_j$ with $j = 1,...,M$;\\
 \hline
\textit{pulses}  & Time-domain waveform of the control pulse sequence.\\
 \hline
 \textit{distorted\_pulses}  & Time-domain waveform of the distorted control pulse sequence (if there are no distortions, the waveform will be identical to the undistorted pulses).\\
  \hline
\textit{expectations}  & The Pauli expectation values 18 or 52 depending on whether one or two qubits (see above). For each state, the order of measurement is: $\sigma_x, \sigma_y, \sigma_z $ applied to the evolved initial states. As the quantum state is evolving in time, the expectations will range within the interval $[1,-1]$. \\
 \hline
 \textit{$V_O$ operator} & The $V_O$ operators corresponding to the three Pauli observables, obtained by averaging the operators $W_O$ over all noise realizations.\\
 
 \hline
  \textit{noise} & Time domain realisations of the relevant noise.\\
   \hline
  \textit{$H_0$}  & The system Hamiltonian $H_0(t)$ for time-step $j$.\\
   \hline
  \textit{$H1$}& The noise Hamiltonian $H_1(t)$ for each noise realization at time-step $j$.\\
  \hline
 \textit{$U_0$}  & The system evolution matrix $U_0(t)$ in the absence of noise at time-step $j$.\\
 \hline
 \textit{$U_I$} & The interaction unitary $U_I(t)$ for each noise realization at time-step $j$.\\
 \hline
 \textit{$V_O$}  & Set of $3 \times 2000$ expectation values (measurements) of the three Pauli observables for all possible states for each noise realization. For each state, the order of measurement is: $\sigma_x, \sigma_y, \sigma_z $ applied to the evolved initial states.\\
 \hline 
 \textit{$E_O$}  & The expectations values (measurements) of the three Pauli observables for all possible states averaged over all noise realizations. For each state, the order of measurement is: $\sigma_x, \sigma_y, \sigma_z $ applied to the evolved initial states.\\
 \hline
 \caption{QDataSet characteristics. The left column identifies each item in the respective QDataSet examples (expressed as keys in the relevant Python dictionary) while the description column describes each item. We elaborate on the description of each item below. }
\label{table:datasetcharacteristics}
\end{longtable}
\end{center}

\newpage
\section{QDataSet Detail \& Background} \label{apnd:detail}

\subsection{State space} Quantum systems are represented by (unit) state vectors within a complex-valued vector (Hilbert) space $\mathcal{H}$ whose dimensionality is determined according to the physics of the problem. The majority of quantum information processing research to date has concentrated on qubits (quantum bits), being arbitrary two-level quantum systems (two-dimensional state spaces) of qubits with arbitrary state vectors with orthonormal bases $\{\kz,\ko\}$:
\begin{align}
\ketpsi = a\kz + b\ko
\label{eqn:qubit}
\end{align}
Qubits are normalised such that they are unit vectors $\braket{\psi | \psi} = 1$ (that is $|a|^2 + |b|^2=1)$, where $a,b \in \C$ are amplitudes for measuring outcomes of $\kz,\ko$ respectively (where $\braket{\psi | \psi'}$ denotes the inner product of quantum states $\ketpsi,\ket{\psi'}$). In density operator formalism, the system is described via a Hermitian positive semi-definite density operator $\rho$ with trace unity acting on the state space of the system (such that if the system is in state $\rho_i$ with probability $p_i$ then $\rho = \sum_i p_i \rho_i$). Density operators are generalised operator-representations of probability distributions over quantum states with particular properties: all their eigenvalues have to be real, non-negative, sum to unity, inheriting the necessary constraints of a probability distribution. In this work, we assume the standard orthonormal computational basis $\{\kz,\ko\}$ such that $\braket{1 | 0}=\braket{0 | 1}=0$ and $\braket{1|1}=\braket{0|0}=1$. Quantum states encode information of interest and use to optimisation problems. They are not directly observable, but rather their structure must be reconstructed from known information about the system. In machine learning contexts, quantum states may be used as inputs, constituent elements in intermediate computations or label (output) data. In the QDataSet, intermediate quantum states at any time step may be reconstructed using the intermediate Hamiltonians and unitaries for each example. The code repository for the QDataSet simulation provides further detail on how quantum state representations are used to generate the QDataSet \cite{perrier_qdataset_2021}. Depending on the machine learning architecture, quantum states will usually be represented as matrices or tensors and may be used as inputs (for example, flattened), label data or as an intermediate input, such as in intermediate layers within a hybrid classical-quantum neural network (see \cite{perrier_quantum_2020, youssry_modeling_2020}). For example, consider the matrix representation of eigenstates of a Pauli $\sigma_z$ operator below:
\begin{align}
\sigma_z = \begin{pmatrix} 
1 & 0 \\
0 & -1
\end{pmatrix} 
\end{align}
In the computational basis, this operator has two eigenstates $\kz,\ko$ for eigenvalues $\lambda=1,-1$:
\begin{align}
\kz =
\begin{pmatrix} 
1 \\
0
\end{pmatrix} \qquad \text{for} \qquad \lambda = 1 \qquad
\ko =
\begin{pmatrix} 
0 \\ 
1
\end{pmatrix} \qquad \text{for} \qquad  \lambda=-1 
\end{align}

where we have adopted the formalism that the $\lambda=1$ eigenstate is represented by $\kz$ and the $\lambda=-1$ eigenstate is represented by $\ko$ (our choise is consistent with Qutip - practitioners should check platforms they are using for the choice of representation). These eigenstates have a density operator representation as:
\begin{align}
\rho_{\lambda=1}=\kz\bz \qquad  \rho_{\lambda=-1} = \ko\bo
\end{align}
with matrix representations:
\begin{align}
\kz\bz = \begin{pmatrix} 
1 & 0 \\
0 & 0
\end{pmatrix}\qquad  \ko\bo = \begin{pmatrix} 
0 & 0 \\
0 & 1
\end{pmatrix}
\end{align}
For machine learning practitioners, one way to think about density operators is associating rows and columns with bra and ket vector representations:
\begin{align}
\rho = a\kz\bz + b\kz\bo + c\ko\bz + d\ko\bo \dot{= }
\begin{blockarray}{ccc}
& \bz & \bo \\
\begin{block}{c (cc)}
\kz \qquad & a & b  \\
\ko \qquad & c & d  \\
\end{block}
\end{blockarray}
\end{align}
where $a,b,c,d \in \C$ are the complex-values amplitudes respective. Given $\rho = \sum_{p_i} \rho_i$, the diagonal elements $a_{n}$ of the density matrix describe the probability $p_n$ of the system residing in state $\rho_n$, that is 
\begin{align}
\rho_{nn} = a_{n}a_n^* = p_n \geq 0
\end{align}
For pure states, the diagonal along the density matrix will only have one non-zero element (i.e. it will be 1) so that $\rho = \rho_i$. A mixed state will have multiple entries along the diagonal such that $ 0 \leq a_n < 1 $. For example, the $\sigma_z$ eigenvectors have the representation:
   \begin{align} 
   \kz\bz \dot{=} 
\begin{blockarray}{ccc}
 & \bz & \bo \\
\begin{block}{c (cc)}
  \kz \qquad & 1 & 0  \\
  \ko \qquad & 0 & 0  \\
\end{block}
\end{blockarray}
\qquad
 \ko\bo \dot{=} 
\begin{blockarray}{ccc}
 & \bz & \bo \\
\begin{block}{c (cc)}
  \kz \qquad & 0 & 0  \\
  \ko \qquad & 0 & 1  \\
\end{block}
\end{blockarray}
 \end{align}
 
   Sometimes the density matrix representation of a state will be equivalent to the outer product of the state, but caution should be applied as this is not generally the case. Translating between the nomenclature and symbolism of quantum information to a more familiar matrix representation used in machine learning assists machine learning researchers to develop their algorithmic architecture. The QDataSet simulation code utilises state space representations of data and operations thereon in order to generate the output contained in the datasets themselves. To recover a quantum state $\ket{\psi(t_j)} = U(t_j)\ket{\psi_0} \approx \prod_i U_i^j \ket{\psi_0}$, one can simply apply the sequence $U_i$ up to $i=j$ (note the order of application is such that $U_j...U_0 \ketpsi$).

\subsection{Operators and evolution}
The Hamiltonian $H(t)$ of a system is primary means of mathematically characterising dynamics of quantum systems. Hamiltonians specify by the data-encoding process by which information is encoded into quantum states along with how the system evolves and how the quantum computation may be controlled. Unitary evolution itself is required to preserve quantum coherence and probability measures of systems (which give rise to the enhanced computational power of quantum systems). Algorithms trained using the QDataSet will usually involve modification to the controllable part of a system Hamiltonian in order to steer a system towards a desired state. This is to be understood more fully in terms of quantum state evolution. 

Closed quantum systems (which we focus on in this paper for simplicity) evolve over time $\Delta t= t_1 - t_0$ via unitary transformations $U(\Delta t)=\mathcal{T}_+\exp\left(-i\int_0^{\Delta t} H(t) dt\right)$ where $\mathcal{T}_+$ is the time-ordering operator (described below). As discussed below, given the difficulties in solving for time dependency, unitaries are typically approximated by time-independent sequences . The evolution of quantum states is characterised by such unitaries operating upon vectors that transforms (transitions) to other states. Intermediate quantum states $\ket{\psi'}$ may be represented as the result of applying unitary operators to initial states $\ketpsi$ such that $\ket{\psi'} = U(t) \ket{\psi_0}$ (or $\rho' = U (t) \rho_0 U(t)^\dagger$). Thus given initial quantum states, quantum state evolution can be represented entirely via operator dynamics and representations. There is a panoply of mathematical formalisms via which to understand operator dynamics, from representation theory, to operator algebras to category theory. From the perspective of a machine learning practitioners, operators will take the form of matrices or tensors in standard programming languages. It is worth noting that the operator acting on a quantum state $\rho$ is a unitary $U(t)$ which is itself (in a closed quantum system) a function (or representation) of the Hamiltonian $H(t)$ governing the system dynamics. In practice unitaries are formed by exponentiating time-independent approximations of Hamiltonians and unitaries. These unitaries represent solutions to the time-dependent Schr{\"o}dinger equation governing evolution:
    \begin{align}
        i\hbar \frac{d\ket{\psi(t)}}{dt}=H(t)\ket{\psi(t)}
    \end{align} 
    where $\hbar$ is set to unity for convenience and $H(t)$ represents the linear Hermitian operator (Hamiltonian) of the closed system. The dynamics of the quantum system are completely described by the Hamiltonian operator acting on the state $\ketpsi$ such that $\ket{\psi(t)} = U(t)\ket{\psi(t=0)}$. In density operator notation, such evolution is represented as $\rho(t) = U(t) \rho(t_0) U(t)^\dagger$. Typically solving the continuous form of the Schr{\"o}dinger equation is intractable or infeasible, so a discretised approximation as a discrete quantum circuit (where each gate $U_i$ is run for a sufficiently small time-step $\Delta$t) is used (e.g. via Trotter-Suzuki decompositions).

\subsection{Composite systems} States $\ketpsi$ in the Hilbert space may be composite systems, described as the tensor product of states spaces of the component physical systems, that is $\ketpsi = \otimes_i \ket{\psi_i}$. We also mention here the importance of open quantum systems where a total system Hamiltonian $H$ can be decomposed as $H = H_S + H_E + H_I$, comprising a closed quantum system Hamiltonian $H_S$, an environment Hamiltonian $H_E$ an interaction Hamiltonian term $H_I$, which is typically how noise is modelled in quantum contexts. Open quantum systems are typically approximated using master equations to capture the dissipative effects of system/environment interaction. The dissipative nature of open quantum systems has parallels with the dissipative characteristics of neural networks (see \cite{schuld_supervised_2018}). Simulated data of open quantum systems can be generated in a number of packages, such as Qutip. For the QDataSet, we made a design decision to directly simulate the effects of coupling to dissipative environmental baths using more elementary Monte Carlo methods. The reason for this is that master equation formalism both requires a number of assumptions on the system (see \cite{wiseman_quantum_2010, johansson_qutip_2013}) which may be difficult to apply to experimental setups. We also chose to manually engineer the effect of baths in order to minimise the theoretical barriers for classical machine learning practitioners using the QDataSet. 

\subsubsection{Measurement} Quantum measurements are framed as sets of measurement operators $\{ M_m\}$, where $m$ indexes the outcome of a measurement (e.g. an energy level or state indicator), i.e. an observable. The probability $p(m)$ of observable $m$ upon measuring $\ketpsi$ is represented by such operators acting on the state such that $p(m) = \braket{\psi | M_m^\dagger M_m |\psi}$ (alternatively, $p(m) = \text{tr}(M_m^\dagger M_m \rho)$) with the post-measurement state $\ket{\psi'}$ given by: 
    \begin{align}
        \ket{\psi'} = \frac{M_m \ketpsi}{\sqrt{\braket{\psi | M_m^\dagger M_m | \psi}}}
        \label{eqn:postmeasurementstate}
    \end{align}
    The set of measurement operators $\sum_m M_m^\dagger M_m = I$, reflecting the probabilistic nature of measurement outcomes. In more advanced treatments, POVM formalism more fully describes the measurement statistics and post-measurement state of the system. There we define a set of positive operators $\{  E_m \}=\{M^\dagger_m M_m\}$ satisfying $\sum_m E_m=\mathbb{I}$ in a way that gives us a complete set of positive operators (such formalism being more general than simply relying on projection operators). As we are interested in probability distributions rather than individual probabilities from a single measurement, we calculate the probability distribution over outcomes via Born rule using the trace $p(E_i) = \trace(\rho E_i)$. We describe measurement procedures for the QDataSet in more detail below.

\subsubsection{Additional concepts}
There are a number of other important concepts for classical machine learning practitioners to be aware of when using quantum datasets such as the QDataSet. We set out some of these: (a) \textit{relative phase}, for a qubit system, amplitudes $a$ and $b$ differ by a relative phase if $a = \exp(i\theta)b, \theta \in \R$ (relative phase is an important concept as the relative phase encodes quantum coherences and is, together with basis encoding, a primary means of encoding data in quantum systems); (b) \textit{entanglement}, composite states (known as EPR or Bell states), may be entangled, meaning that their measurement outcomes are necessarily correlated. For a two-qubit state: 
\begin{align}
    \ketpsi = \bellzz
\end{align}
a measurement outcome of $0$ on the first qubit necessarily means that a measurement of the second qubit will result in the post-measurement state $\kz$ also i.e. the measurement statistics of each qubit correlate. States that are entangled cannot be written as tensor products of component states i.e. $\ketpsi \neq \ket{\psi_1}\ket{\psi_2}$; (c) \textit{expectation}, expectation values of an operator $A$ (e.g. a measurement) can be written as $E(A) = \text{tr}(\rho A)$ (see below); (d) \textit{mixed} and \textit{pure} states, if the state of a quantum system is known exactly $\ketpsi$, i.e. where $\psi = \ketpsi\bra{\psi}$ then it is denoted as a \textit{pure states}, while where there is (epistemic) uncertainty about its state, it is a mixed state i.e. $\rho = \sum_i p_i \rho_i$ where $\text{tr}(\rho^2)< 1$ (as all $p_i < 1$); (e) \textit{commutativity}, multiple measurements are performed on a system, the outcome will be order-dependent if the measurement operators corresponding to those measurements do not commute i.e if $[A,B]\neq 0$. This is distinct from the classical case; and (f) \textit{no cloning}, quantum data cannot be copied, fundamentally distinguishing it from classical data. There are a range of other aspects of quantum systems that are relevant to the use of machine learning algorithms for solving optimisation problems which we pass over but which are relevant to research programmes using such algorithms, including the role of error-correcting codes (designed to limit or self-correct errors to achieve fault-tolerant quantum computing). While not the focus of the QDataSet, it is worth noting for machine learning practitioners that a distinction is usually drawn in the quantum information literature between \textit{physical} and \textit{logical} qubits. A physical qubit is a two-level physical quantum system. A logical qubit is itself an abstraction from a collection of physical qubits which in aggregate behave according to qubit parameters \cite{shaw_encoding_2008}. The QDataSet is generated on the assumption that each qubit is a logical qubit (which may or not equate to a single corresponding physical qubit). For more complex treatments (involving a multitude of physical qubits) in quantum control or quantum error correction, the underlying simulation code may be adapted accordingly.

\subsection{Quantum metrics}
Metrics play a central technical role in classical machine learning, fundamentally being the basis upon which machine learning algorithms update, via techniques such as backpropagation. Metrics for quantum information processing are related but distinct from their classical counterparts and understanding these differences is important for researchers applying classical machine learning algorithms to solve problems involving quantum data. As is commonplace within machine learning, chosen metrics will differ depending on the objectives, optimisation strategies and datasets. For a classical bit string, there are a variety of classical information distance metrics used in general \cite{nielsen_quantum_2010}. In more theoretical and advanced treatments, available metrics will depend upon the underlying structure of the problem (e.g. topology) (see \cite{watrous_theory_2018} for a comprehensive discussion). Metrics used will depend also upon whether quantum states or operators are used as the comparators, though one can relatively easily translate between operator and state metrics. We outline a number of commonly used quantum metrics below and discuss their implementation in classical contexts, such as in loss functions. Note below we take license with the term metric as certain measures below, such as quantum relative entropy do not (as with their classical counterparts) strictly constitute metrics as such.     
\begin{enumerate}
    \item \textit{Hamming distance}, the number of places at which two bit strings are unequal. Hamming distance is important in error-correcting contexts and quantum communication \cite{doriguello_quantum_2019}.
    \item \textit{Trace distance} or $L1$-\textit{Kolmogorov distance}. For two probability distributions $\{p_x \},\{q_x \}$ we can construct the metric $D(p_x,q_x)=\frac{1}{2}\sum_x|p_x - q_x|$ . In the quantum setting, for states represented by density matrices $\rho,\sigma$ \cite{nielsen_quantum_2010}, their trace distance can be calculated as:
\begin{align}
    D(\rho,\sigma) = \frac{1}{2} \text{tr}|\rho - \sigma|
\end{align}
where $|\rho| = \sqrt{\rho^\dagger \rho}$ is taken as the positive square root. Trace distance is a metric which is preserved under unitary transformations, thus is a widely used similarity metric in quantum information.
       \item \textit{Fidelity} is another  common metric by which to assess state or operator similarity. Fidelity is given by $F(\rho,\sigma) = \text{tr}\sqrt{\rho^{1/2}\sigma \rho^{1/2}}$. It is among the most important metrics in quantum computing, being the measure by which quantum states or operators are measured. Fidelity can be interpreted as a metric by calculating the angle $\zeta=\arccos F(\rho,\sigma)$. Fidelity and trace distance are  via $D(\rho,\sigma) = \sqrt{1 - F(\rho,\sigma)^2}$.
    \item \textit{quantum relative entropy}, is the quantum analogue of Shannon entropy. It is found given by $S(\rho) = -\text{tr}(\rho \log \rho)$ The quantum analogue of (binary) cross-entropy is in turn given by:
    \begin{align}
        S(\rho||\sigma) = \text{tr}(\rho \log \rho) - \text{tr}(\rho \log \sigma)
    \end{align} These measures provide a further basis for comparing for the output of algorithms to labelled data during training. 
\end{enumerate}
For most machine learning practitioners using the QDataSet, the entry point will be the application of known classical machine learning metrics. More advanced uses of the QDataSet may utilise quantum-specific metrics directly, for example, via reconstruction of quantum states from measurement statistics. Some use cases combine the use of classical and quantum metrics. For example, \cite{perrier_quantum_2020, youssry_characterization_2020} combine average operator fidelity with standard mean-squared error (MSE) into a measure denoted as `batch fidelity'. In those examples, the objective in question is to train a greybox algorithm to model certain control pulses needed to synthesise target unitaries. The algorithms learn the particular control pulses which are applied to generators. While it is the extraction of control pulses which are of interest to experimentalists, the final output of the algorithm is a sequence of fidelities where the fidelity of generated (synthesised) unitaries is compared against the target (label) unitaries $U_T$. This sequence of fidelities is then compared against a vector of ones with the loss function set to minimise the MSE (distance) between the fidelity sequence and the label vector. In doing so, the algorithms are effectively trained to maximise fidelity (as fidelities $\approx 1$ are desirable) yet do so using a hybrid approach. The QDataSet has been generated such that a combination of classical, quantum and hybrid metrics of divergence measures may be used in the training process.

\subsection{Encoding data in quantum systems}
Most quantum information optimisation problems involve information encoded in quantum systems, either by construction in an experiment involving quantum systems themselves, or via encoding exogenous or classical information into quantum systems (such as qubits) in order to leverage the benefits of quantum computation. Both approaches involve the input into quantum states in a process known as \textit{state preparation}. The way in which data is encoded in quantum systems affects the performance and expressiveness many quantum algorithms [\textbf{Schuld} 2021] . Information is usually encoded using one of four standard encoding methods including \cite{schuld_supervised_2018}: (a) basis encoding, (b) amplitude encoding, (c) qsample encoding and (d) dynamic encoding. The first of these, \textit{basis encoding}, is a technique that encodes classical information into quantum basis states. Usually, such procedures involve transformation of data into classical binary bit-strings $(x_1,...,x_d), b_i \in \{0,1\}$ then mapping each bit string to the quantum basis state of a set of qubits of a composite system. For example, for $x \in \R^N$, say a set of decimals, is converted into a $d$-dimensional bit string (e.g. $0.1\to 00001..., -0.6 \to 11001..$) suitably normalised such that $x = \sum_k^d (1/2^k)x_k$. The sequence $x$ is given a representation via $\ketpsi = \ket{000001\,11001}$ (see \cite{schuld_supervised_2018}). \textit{Amplitude encoding} associates normalised classical information e.g. for an $n$-qubit system (with $2^n$ different possible (basis) states $\ket{j}$), a normalised classical sequence $x\in \C^{2^n}, \sum_k |x_k|^2=1$ (possibly with only real parts) with quantum amplitudes $x=(x_1,...,x_{2^n})$ can be encoded as $\ket{\psi_x}=\sum_j^{2^n} x_j \ket{j}$. Other examples of sample-based encoding (e.g. \textit{Qsample} and \textit{dynamic} encoding are also relevant but not addressed here). From a classical machine learning perspective, such encoding regimes also enable both features and labels to be encoded into quantum systems.

\subsection{QDataset evolution, noise and measruement}
\subsubsection{Evolution, Hamiltonians and control}
The QDataSet is constructed on the basis of a typical quantum control \cite{sachkov_control_2009, dalessandro_introduction_2007} problem, where it is assumed there exist a set of controls (such as pulses or voltages) with which the quantum system may be controlled or steered towards a target state (or unitary). In density matrix formalism, the Schr{\"o}dinger equation is:.
    \begin{align}
        i\hbar \frac{d \rho}{dt} = [H(t), \rho(t)],
    \end{align}
    where the commutator $[A,B] = AB-BA$, $\hbar$ is Planck's constant (we can always choose the units such that $\hbar=1$ which will be the convention in this paper), and $H(t)$ is a Hermitian operator called the Hamiltonian. In physical systems, it corresponds to the total energy (sum of kinetic and potential energies) of the system under consideration. For a closed system (i.e. a noiseless isolated system with no interaction with the surrounding environment), it can be expressed in the general form
    \begin{align}
        H(t) = H_0(t) \doteq H_d(t) + H_{\text{ctrl}}(t).
    \end{align}
    $H_d(t)$ is called the drifting Hamiltonian and corresponds to the natural evolution of the system in the absence of any control. The second term $H_{\text{ctrl}}(t)$ is called the control Hamiltonian and corresponds to the controlled external forces we apply to the system (such as electromagnetic pulses applied to an atom, or a magnetic field applied to an electron). 
        The solution of the evolution equation at time $t=T$ is given by:
    \begin{align}
        \rho(T) = U(T) \rho(0) U^{\dagger}(T),
    \end{align}
    where $\rho(0)$ is the initial state of the system, the unitary evolution matrix $U(t)$ is given by:
    \begin{align}
        U(T) &= \mathcal{T}_{+} e^{-i\int_0^T H(t) dt}\\
             &\doteq \lim_{\Delta t\to 0} e^{-i H(T) \Delta t} \cdots e^{-i H(3\Delta t) \Delta t} e^{-i H(2\Delta t) \Delta t} e^{-i H(\Delta t) \Delta t}
    \end{align}
    The symbol $\mathcal{T}_+$ is called a time-ordering operator.  The time-ordering is needed because in general the Hamiltonian is time-dependent and does not commute at different time instants (i.e. $[H(t_i), H(t_j)]\neq 0$). The second line is a time-independent approximation of the time-dependent form based on a Suzuki-Trotter decomposition \cite{suzuki_quantum_1993, childs_theory_2019}:
    \begin{align}
        e^{A + B} = \lim_{n\to \infty}\left( e^{A/n}e^{B/n}\right)^n
    \end{align}
    which under certain conditions allows the time-varying Hamiltonian to be approximated by a piece-wise constant Hamiltonian. In this case where the time interval $[0,T]$ is divided into equal segments of length $\Delta t$. As a special case if the Hamiltonian commute at different time instants then the evolution matrix can be simplified to: 
    \begin{align}
        U(T) &= e^{-i\int_0^T H(t) dt},
    \end{align}
    which also reduces to: 
    \begin{align}
        U(T) &= e^{-i H(T) T},
    \end{align}
    in the case of a time-independent Hamiltonian.
    
    On the other hand when the system is open (i.e. uncontrollable interactions with the environment), then the Hamiltonian takes the form:
    \begin{align}
        H(t) = H_0(t) + H_1(t) \doteq H_d(t) + H_{\text{ctrl}}(t) + H_{SE}(t) + H_E(t).
    \end{align}
    $H_0(t)$ is defined as before. The new term $H_1(t)$ now consists of two terms: $H_{SE}(t)$ represents an interaction term with the environment, while $H_E(t)$ represents the free evolution of the environment in the absence of the system. In this case, the  Hamiltonian controls the dynamics of both the system and environment combined in a highly non-trivial way. In other words, the state becomes the joint state between the system and environment. The combined system and environment then become closed. Modelling such open quantum systems is complex and challenging and is typically undertaken using a variety of stochastic master equations \cite{wiseman_quantum_2010} or sophisticated noise spectroscopy. As detailed below, the QDataSet contains a variety of noise realisations for one and two qubit systems together with details of a recent novel operator \cite{youssry_beyond_2020} for characterising noise in quantum systems.

\subsubsection{Hamiltonians: drift, control, noise}
The QDataSet comprises datasets for 1- and 2- qubits systems evolving in the presence and absence of noise. The canonical forms of Hamiltonian in the QDataSet are that given in \cite{youssry_characterization_2020}. In that work, a limited set of single-qubit systems subject to external environmental noise (baths) was used as input training data for a novel greybox machine learning alternative method for characterising noise (`beyond' conventional spectroscopic techniques). In this paper, the underlying simulation was modified to generate a greater variety of qubit-noise examples for the single qubit case. The simulation was then extended beyond that in \cite{youssry_characterization_2020} to generate examples for the two-qubit case (in the presence or absence of noise). As discussed above, the evolution of closed and open quantum systems is described by Hamiltonian dynamics, which encode time-dependent functions into operators which are the generators of time-translations (operators) acting on quantum states. The general form of the Hamiltonian for the QDataSet is given by:
\begin{align}
    H(t)&= H_{\text{drift}}(t) + H_{\text{ctrl}}(t) + H_1(t)
\end{align}
The Hamiltonian comprises three elements: (i) a drift Hamiltonian $H_d(t)$, encoding the endogenous evolution of the quantum system; (ii)  a control Hamiltonian $H_{\text{ctrl}}(t)$, encoding evolution due to the application of classical controls which may be applied to the quantum system to steer its evolution; and (iii) and an interaction (noise) Hamiltonian $H_1(t)$, encoding the effects of coupling of the quantum system to its environment, such as a decohering noise source (a bath). We express the Hamiltonians in the Pauli operator basis which forms a complete basis for our one- and two-qubit systems. Our control functions are represneted as $f_\alpha(t)$ for $\alpha = x,y,z$ where the subscript indicates which generator the control applies to. Concretely, for example, $\sigma_z$ control is denoted $f_z(t) \sigma_z$. In general, continuous time-dependent control formulations are difficult - or infeasible - to solve analytically, where solving here means finding a suitable representation of the control unitary: 
\begin{align}
    U_{\text{ctrl}} = \mathcal{T}_{+} e^{-i\int_0^T f_\alpha \sigma_\alpha/2 dt}
\end{align}
The control functions act on the The simplest control functional form is fixed amplitude control \cite{schattler_geometric_2012} or what is also described as a \textit{square pulse}, where a constant energy (expressed as amplitudes) is expressed for a discrete time-step $\Delta t$. Most controls are usually [classical] i.e. $f_\alpha(t) \in \R$. Other control waveforms include Gaussian pulses which are characterised by [\textbf{insert}]. Moreover, quantum control in the QDataSet context has two primary imperatives. The first is the use of control in closed noise-free idealised quantum systems where the objective is the use of controls to steer the quantum system to some desired objective state. This is equivalent to the synthesis of quantum circuits (sequences of quantum gates) from the identity $I$ to a target unitary $U_T$. The second is the use of controls in the presence of noise, where the quantum system is coupled to an environment that potentially decoheres the system. In this second case, ideally the controls are tailored to neutralise the effect of noise on the evolution of the quantum system - a process typically described by dynamic decoupling \cite{gupta_machine_2018, mavadia_prediction_2017} (see for example Hahn echo or other examples). Crafting suitable controls to mitigate noise effects is challenging. One must properly time and select appropriate amplitudes for the application of controls to counteract decohering interference. Typically, it requires information about the noise spectrum itself, obtained using techniques from quantum noise spectroscopy \cite{wiseman_quantum_2010}. It also requires an understanding of how control and noise signals convolve in the context of quantum evolution (see [ref]). Dealing with noise is a central imperative of quantum information processing and the engineering of quantum systems where the aim is to preserve and extend coherence times of quantum information). The simulations in the QDataSet are based upon an alternative technique for quantum control in the presence of a variety of noise \cite{youssry_characterization_2020}, where a greybox neural network is used to learn only those characteristics of the noise spectrum relevant to the application of controls (a comparatively simpler problem than seeking to determine the entire spectrum). For machine learning practitioners, the QDataSet thus provides a useful way to seek to apply advanced classical machine learning techniques to the challenging but important problem. 
\subsubsection{One- and two-qubit Hamiltonians}
We begin by describing the single-qubit Hamiltonian and then move to an exposition of the two-qubit case. For the single-qubit system, the drifting Hamiltonian is fixed in the form:
\begin{align}
    H_d(t) = H_d =  \frac{1}{2} \Omega \sigma_z.
\end{align}
The $\Omega$ term represents the energy gap of the quantum system (the difference in energy between, for example, the ground and excited state of the qubit, recalling qubits are characterised by having two distinct quantum states). The single-qubit drift Hamiltonian for the QDataSet is time-independent for simplicity, though in realistic cases it will contain a time-dependent component. For the single-qubit control and noise Hamiltonians we have two cases based upon the concept of which axes controls and noise are applied. Recall we can represent a single qubit system on a Bloch sphere, with axes corresponding to the expectations of each Pauli operator and where operations of each Pauli operator constitute rotations about the respective axis. Our controls are control functions, mostly time-dependent, that apply to each Pauli operator (generator). They act to affect the amplitude over time of rotations about the respective Pauli axes. More detailed treatments of noise in quantum systems and quantum control contexts can be found in \cite{wiseman_quantum_2010}.

As discussed above, the functional form of the control functions $f_\alpha(t)$ varies. We select both square pulses and Gaussian pulses as the form (see below). Each different noise function $B_\alpha(t)$ is parameterised differently depending on various assumptions that are more specifically detailed in \cite{youssry_characterization_2020} and summarised below.  Noise and control functions are applied to different qubit axes in the single-qubit and two-qubit cases. For a single qubit, we first have single-axis control along $x$-direction:
\begin{align}
    H_{\text{ctrl}}(t) &= \frac{1}{2} f_x(t) \sigma_x 
\end{align}
with the noise (interaction) Hamiltonian $H_1(t)$ along $z$-direction (the quantification axis):
5
\begin{align}
    H_1(t) &= \frac{1}{2} \beta_z(t) \sigma_z
\end{align}
Here the function $\beta_z(t)$ (a classical noise function $\beta(t)$ applied along the $z$-axis) may take a variety of forms depending on how the noise was generated (see below for a discussion of noise profiles e.g. N1-N6). It should be noted (for researchers unfamiliar with noise) noise rarely has a functional form and is itself difficult to characterise (so $\beta(t)$ should not be thought of as a simple function). For the second case, we implement multi-axis control along $x$- and $y$- directions and noise along $x$- and $z$-directions in the form:
\begin{align}
    H_{\text{ctrl}}(t) &= \frac{1}{2} f_x(t) \sigma_x + \frac{1}{2} f_y(t) \sigma_y \\
    H_1(t) &= \frac{1}{2} \beta_x(t) \sigma_x + \frac{1}{2} \beta_z(t) \sigma_z.
\end{align}
Noiseless evolution may be recovered by choosing $H_1(t)=0$.
For the 2-qubit system, we chose the drifting Hamiltonian in the form:
\begin{align}
    H_d(t) = \frac{1}{2} \Omega_1 \sigma_z \otimes \sigma_0 + \frac{1}{2} \Omega_2 \sigma_0 \otimes \sigma_z.
\end{align}
For the control Hamiltonians, we also have two cases. The first one is local control along the $x$-axis of each individual qubit, akin to the single-qubit case each. In the notation, $f_{1\alpha}(t)$ indicates that the control function is applied to, in this case, the second qubit, while the first qubit remains unaffected (denoted by the `1' in the subscript and by the identity operator $\sigma_0$). We also introduce an \textit{interacting control}. This is a control that acts simultaneously on the $x$-axis of each qubit, denoted by $f_{xx}(t)$:  
\begin{align}
     H_{\text{ctrl}}(t) = \frac{1}{2} f_{x1}(t) \sigma_x \otimes \sigma_0 + \frac{1}{2} f_{1x} \sigma_0 \otimes \sigma_x + f_{xx}(t) \sigma_x \otimes \sigma_x.
\end{align}
The second two-qubit case is for local-control along the $x-$axis of each qubit only and is represented as:
\begin{align}
     H_{\text{ctrl}}(t) = \frac{1}{2} f_{x1}(t) \sigma_x \otimes \sigma_0 + \frac{1}{2} f_{1x} \sigma_0 \otimes \sigma_x,
\end{align}
For the noise, we fix the Hamiltonian to be along the $z$-axis of both qubits, in the form: 
\begin{align}
    H_{1}(t) &= \frac{1}{2} \beta_{z1}(t) \sigma_z \otimes \sigma_0 + \frac{1}{2} \beta_{1z} \sigma_0 \otimes \sigma_z.
\end{align}
Notice, that for the case of local-only control and noiseless evolution, this will correspond to two completely-independent qubits and thus we do not include this case, as it is already covered by the single-qubit datasets.
\\
To summarise, the QDataSet includes four categories for the datasets set-out in Table \ref{tab:cats}. The first two categories are for 1-qubit systems, the first is single axis control and noise, while the second is multi-axis control and noise. The third and fourth categories are 2-qubit systems with local-only control or with an additional interacting control together with noise. 

\begin{table}[h]
    \centering
    \begin{tabular}{|c|c|c|c|c|}
    \hline
         Category &  Qubits & Drift & Control & Noise \\
         \hline
         1 & 1 & $(z)$ & $(x)$ & $(z)$ \\
         \hline
         2 & 1 & $(z)$ & $(x,y)$ & $(x,z)$ \\
         \hline
         3 & 2 & $(z1, 1z)$ & $(x1, 1x)$ & $(z1, 1z)$ \\
         \hline
         4 & 2 & $(z1, 1z)$ & $(x1, 1x, xx)$ & $(z1, 1z)$ \\
         \hline
    \end{tabular}
    \caption{The general categorization of the provided datasets.}
    \label{tab:cats}
\end{table}

\subsubsection{Control}
The control pulse sequences in the QDataSet consist of two types of waveforms. The first is a train of Gaussian pulses, and the other is a train of square pulses, both of which are very common in actual experiments. Square pulses are the simplest waveforms, consisting of a constant amplitude $A_k$ applied for a specific time interval $\Delta t_k$:
\begin{align}
    f(\Delta t_k) = f = A_k \Delta t_k
\end{align}
where $k$ runs over the total number of time-steps in the sequence. The three parameters of such square pulses are the amplitude $A_k$, the position in the sequence $k$ and the time duration over which the pulse is applied $\Delta t$. In the QDataSet, the pulse parameters are stored in a sequence of vectors $\{a_n \}$. Each vector $a_n$ is of dimension $r$ parameters of each pulse (e.g. the Gaussian pulse vectors store the amplitude, mean and variance, the square pulse vectors store pulse position, time interval and amplitude), enabling reconstruction of each pulse from those parameters if desired. For simplicity, we assume constant time intervals such that $\Delta t_k = \Delta t$. The Gaussian waveform can be expressed as: 
\begin{align}
    f(t) = \sum_{k=1}^{n} A_k e^{-(t-\mu_k)^2 /2 \sigma_k^2}.
\end{align}
A sequence of $n$ Gaussian pulses is defined by $3n$ parameters, being (for each pulse): (i) the amplitude $A_k$, (ii) the mean $\mu_k$ and (iii) the variance $\sigma_k$ of the pulse sequence.
The amplitudes for both Gaussian and square pulses are chosen uniformly at random from the interval $[A_{\text{min}}, A_{\text{max}}]$, the standard deviation for all Gaussian pulses in the train is fixed to $\sigma_k = \sigma$, and the means are chosen randomly such that there is no overlap between the pulses in the train. The square pulses are generated similarly to the Gaussian pulses, with the pulse width chosen to be equal to $6\sigma$. The pulse sequences can be represented in the time or frequency domains \cite{bandrauk_quantum_2003}. The QDataSet pulse sequences are represented using the time-domain as it has been found to be more efficient feature for machine learning algorithms \cite{youssry_characterization_2020}.

As discussed in \cite{youssry_characterization_2020}, the choice of structure and characteristics of quantum datasets depends upon the particular objectives and uses cases in question, the laboratory quantum control parameters and experimental limitations. Training datasets in machine learning should ideally be structured so as to enhance the generalisability. In the language of statistical learning theory, datasets should be chosen so as to minimise the empirical risk associated with candidate sets of classifiers \cite{vapnik_nature_1995, hastie_elements_2013}. In a quantum control context, this will include understanding for example the types of controls available to researchers or in experiments, often voltage or (microwave) pulse-based \cite{hollenberg_charge-based_2004}. The temporal spacing and amplitude of each pulse in a sequence of controls applied during an experiment may vary by design or to some degree uncontrollably. Pulse waveforms can also differ. For example, the simplest pulse waveform is a constant-amplitude pulse applied for some time $\Delta t$ \cite{khaneja_optimal_2005}. Such pulses are characterised by for example a single parameter, being the amplitude of the waveform applied to the quantum system (this manifests as we discuss below as an amplitude applied to the algebraic generators of unitary evolution (see \cite{dalessandro_introduction_2007, perrier_quantum_2020} for an example). Other models of pulses (such as Gaussian) are more complex and require more sophisticated parametrisation and encoding with machine learning architectures in order to simulate. More detail on such considerations and the particular pulse characteristics in the QDataSet are set-out in Table (\ref{table:datasetproperties}).

\subsection{QDataSet Noise}

\subsubsection{Noise characteristics}
Noise applicable to quantum systems is generally classified as either \textit{classical} or \textit{quantum} \cite{paz-silva_extending_2019}. Classical noise is represented typically as a stochastic process \cite{wiseman_quantum_2010} and can include, for example (i) slow noise which is pseudo-static and not varying much over the characteristic time scale of the quantum system and (ii) fast or `white' noise with a high frequency relative to the charateristic frequencies (energy scales) of the system \cite{wardrop_exchange-based_2014}. The effect of quantum noise quantum systems of noise is usually characterised into two forms. The first is dephasing ($T_2$) noise, which characteristically causes quantum systems to decohere, thus destroying or degrading quantum information encoded within qubits. Such noise is usually characterised as an operator acting transverse to the quantisation axis of chosen angular momentum. 

What this means in practice is usefully construed as follows using a Bloch sphere. Once an orientation ($x,y,z$-axes) is chosen, one is effectively choosing a choice of basis i.e. the basis of a typical qubit $\ketpsi = a\kz + b \ko$ is the basis of eigenstates of the $\sigma_z$ operator. When noise acts along the $z$-axis (i.e. is associated to the $\sigma_z$ operator), then it has the potential to (if the energy of the noise is sufficient) shift the energy state in which the quantum system is in, represented by a `flip' in the basis from $\kz$ to $\ko$ for example. This type of noise is $T_1$ noise. By contrast, noise may act along $x$- and $y$-axes of a qubit, which is represented as being associated with the $\sigma_x$ and $\sigma_y$ operators. These axes are `transverse' to the quantisation axis. Noise along these axis has the effect of dephasing a qubit, thus affecting the coherences encoded in the relative phases of the qubit. Such noise is denoted $T_2$ noise. Open quantum systems' research and understanding noise in quantum systems is a vast and highly specialised topic. As we describe below, the QDataSet adopts the novel approach outlined in \cite{youssry_characterization_2020} where, rather than seeking to fully characterise noise spectra, the only the information about noise relevant to the application of controls (to dampen noise) is sought. Such information is encoded in the $V_O$ operator, which is an expectation that encodes the influence of noise on the quantum system. In a quantum control problem using the QDataSet samples containing noise, for example, the objective would then be to select controls that neutralise such effects.

\subsubsection{QDataSet noise profiles}
In this paper, we chose a set of noise profiles with different statistical properties. The selected noise profiles have been chosen to emulate commonplace types of noise in experimental settings. Doing so improves the utility of algorithms trained using the QDataSet for application in actual experimental and laboratory settings. While engineers and experimentalists across quantum disciplines will usually be familiar with the theoretical and practical aspects of noise in quantum systems, many machine learning and other researchers to whom the QDataSet is directed will not. To assist machine learning practitioners whom may not be familiar with elementary features of noise, it is useful to understand a number of conceptual classifications related to noise used in the QDataSet as follows: (i) power spectral density (which describes the distribution of the noise signal over frequency); (ii) white noise (usually high-frequency noise with a flat frequency); (iii) colored noise, a stochastic process where values are correlated spatially or temporally; (iv) autocorrelated stochasticity, which describes where the noise waveform characteristics are biased by tending to be short (blue) or long (red) as distinct from unautocorrelated noise, where waveforms are relatively uniformly distributed across wavelengths; and (v) stationary noise (a waveform with  a constant time period) and non-stationary noise (a waveform with a varying time period). See literature on noise in signal processing for more detail. The noise realizations are generated in time domain following one of these profile listed as follows (see \cite{youssry_characterization_2020} for specific functional forms):
\begin{itemize}
\item N0: this is the noiseless case (indicated in the QDataSet parameters as set out in Table (\ref{table:datasetproperties});
    \item N1: the noise $\beta(t)$ is described by its power spectral density (PSD) $S_1(f)$, a form of $1/f$ noise with a Gaussian bump;
    \item N2: here $\beta(t)$ is stationary Gaussian colored noise described by its autocorrelation matrix; chosen such that it is colored, Gaussian and stationary (typically lower frequency) and is produced via convolving Gaussian white noise with a deterministic signal;
    \item N3: here the noise $\beta(t)$ is non-stationary Gaussian colored noise, again described by its autocorrelation matrix which is chosen such that it is colored , Gaussian and non-stationary. The noise is simulated via multiplication of a deterministic time-domain signal with stationary noise;
    \item N4: in this case, the noise $\beta(t)$ is described by its autocorrelation matrix chosen such that it is colored, non-Gaussian and non-stationary. The non-Gaussianity of the noise is achieved via squaring the Gaussian noise so as to achieve requisite non-linearities;
    \item N5: a noise described by its power spectral density (PSD) $S_5(f)$, differing from N1 only via the location of the Gaussian bump; and
    \item N6: this profile is to model a noise source that is correlated to one of the other five sources (N1 - N5) through a squaring operation. If the $\beta(t)$ is the realization of one of the five profiles, N6 will have realizations of the form $\beta^2(t)$. This profile is used for multi-axis and multi-qubit systems. 
\end{itemize}
The N1 and N5 profiles can be generated following the method described in \cite{youssry_characterization_2020}. Regarding the other profiles, any standard numerical package can generate white Gaussian stationary noise. The QDataSet noise realisations were encoded using the Numpy package of Python. We deliberately did so in order to avoid various assumptions used in common quantum programming packages, such as Qutip. To add coloring, we convolve the time-domain samples of the noise with some signal. To generate non-stationarity, we multiply the time-domain samples by a some signal. finally, to generate non-Gaussianity, we start with a Gaussian noise and apply non-linear transformation such as squaring. The last noise profile is used to model the case of two noise sources that are correlated with each other. In this case we generate the first one using any of the profiles N1-N5, and the other source is completely determined.

\subsubsection{Distortion}
In physical experiments, there control pulses are physical signals (such as microwave pulses), which propagate along cables and get processed by different devices. This introduces distortions which cannot be avoided in any real devices. However, by properly engineering the systems, the effects of these distortions can be minimized and/or compensated for in the design. In this paper, we used a linear-time invariant system to model distortions of the control pulses, and the same filter is used for all datasets. We chose a Chebychev analog filter \cite{popescu_chebyshev_2016}, the undistorted control signal is the input, and thus the filter output is the distorted signal. Table (\ref{tab:params}) sets out a summary of key parameters.

% \begin{wrapfig}{}
% \begin{wraptable}%{r} %{5.5cm}

% \end{wraptable}

\subsubsection{QDataSet noise operators}
The QDataSet comprises datasets for single-qubit and two-qubit systems. In developing the datasets, we have assumed that the environment affecting the qubit is classical and stochastic, namely that $H_1(t)$ will be a stochastic term that acts directly on the the system. The stochasticity of $H_1(t)$ means that the expectation of any observable measured experimentally will be given as:
\begin{align}
    \braket{O}_c = \braket{\trace{(\rho(T) O)}}_c,    
\end{align}
where $O$ represents the measurement operator corresponding to the observable of interest (e.g. $O \dot{=} M_m$ in notation above) and the $\braket{\cdot}_c$ is a classical expectation over the distribution of the noise realizations. It can then be shown \cite{youssry_beyond_2020} that this can be expressed in term of the initial state $\rho(0)$, and the evolution fixed over the time interval $[0,T]$ as
\begin{align}
    \braket{O(T)} = \trace{\left(V_O(T) U_0^{\dagger}(T) \rho(0) U_0(T) O\right)}
    \label{eqn:voeqn}
\end{align}
where $U_0(T) = \mathcal{T}_{+} e^{-i\int_0^T H_0(t) dt}$ is the evolution matrix in the absence of noise, and 
\begin{align}
    V_O(T) &= \braket{W_O(T)}_c\\
        &\doteq \braket{O^{-1}\tilde{U}_I^{\dagger}(T) O \tilde{U}_I(T)}_c.
\end{align}
is a novel noise operator introduced in \cite{youssry_characterization_2020} which characterises the expectation of noise relevant to synthesising counteracting control pulses. We encapsulate the full quantum evolution via the operator $W_O(T)$. Note that $V_O$ is formed via the partial tracing out of the effect of the noise (bath) and its interaction with the control pulses, so encodes only those features of importance or relevance to impact of noise (not the full noise spectrum). Importantly, the use of the $V_O$ operator is designed to allow information about noise and observables to be separated (a hallmark of dynamic decoupling approaches).  
The modified interaction unitary $\tilde{U}_I(T)$ is defined such that
\begin{align}
    U(T) = \tilde{U}_I(T) U_0(T),
\end{align}
where $U(T)= \mathcal{T}_{+} e^{-i\int_0^T H(t) dt} $ is the full evolution matrix. This contrasts the conventional definition of the interaction unitary which takes the form $U(T) = U_0(T) U_I(T) $. The $V_O$ operator is used in the simulation code for the QDataSet to generate [for each time step and for each noise type] [expectation] values. Ideally, in a noise-free scenario, those expectations should tend to zero (representative of the absence of noise). The idea for including such noise operators is that this data can then be input into machine learning models to assist the algorithms to learn appropriate, for example, pulse sequences or controls that send $V_O \to I$ (neutralising the noise). 

A detailed explanation and example of the use of the $V_O$ operator is provided in \cite{youssry_characterization_2020}. For machine learning practitioners, the operator $V_O$ may, for example, be used in an algorithm that seeks to negate the effect of $V_O$ i.e. learning the sequence of control pulses necessary such that by applying learnt sequences of pulses, $V_O$ effectively becomes $I$. The utility of this approach is that full noise spectroscopy is not required.

\subsection{QDataset Measurement}

\subsubsection{QDataSet POVMs}
As discussed above, quantum measurements are inherently probabilistic. Measurement of quantum systems yields an underlying probability distribution over the possible measurement outcomes (observables) $m_i$ of the system which are in turn determined by the state of the system and the measurement process.  There are several ways to describe quantum measurements mathematically. The most common way (which will be used in this paper), is called protective measurements. In this case, an observable $O$ is described mathematically by a Hermitian operator. The eigendecomposition of the operator can be expressed in the form $O= \sum_{m} m P_m$, where $m$ are the eigenvalues, and $P_m$ are the associated projectors into the corresponding eigenspace. The projectors $P_m$ must satisfy that $P_m^2 = P_m$, and that $\sum_m P_m = I$ (the identity operator), to ensure we get a correct distribution for the outcomes. In more sophisticated treatments, the operators $O$ belong to a POVM described above which partition the Hilbert space $\mathcal{H}$ into distinct projective subspaces $\mathcal{H}_m$ associated with each POVM operator $O$. The probability of measuring an observable is given by:
\begin{align}
    \Pr(m) = \trace(\rho P_m),
\end{align}
for a system in the state $\rho$. The expectation value of the observable is given by:
\begin{align}
    \braket{O}= \trace{(\rho O)} = \trace{\left(\rho \sum_m m P_m\right)} = \sum_m m \Pr(m).
\end{align}
As detailed below, the QDataSet contains measurement outcomes for a variety of noiseless and noisy systems. The POVM chosen is the set of Pauli operators for one and two-qubit systems. The measurement operators chosen are the Pauli operators described below and the QDataSet contains the expectation values for each Pauli measurement operator. In a classical machine learning context, these measurement statistics form training data labels in optimisation problems, such as designing algorithms that can efficiently sequence control pulses in order to efficiently (time-minimally) synthesise a target state or unitary (and thus undertake a quantum computation) of interest.

\subsubsection{Pauli matrices}
The POVM set for the QDataSet is the set of Pauli operators. Pauli operators are important operators in quantum information processing involving qubit systems in part because such qubit systems can be usefully decomposed into a Pauli operator basis via the Pauli matrices:
\begin{align}
    \sigma_x = \begin{pmatrix}0 & 1 \\ 1 & 0\end{pmatrix}, \sigma_y = \begin{pmatrix}0 & -i \\ i & 0\end{pmatrix}, \sigma_z &= \begin{pmatrix}1 & 0 \\ 0 & -1\end{pmatrix}
\end{align}
together with the identity (denoted $\sigma_0$). Pauli operators are Hermitian (with eigenvalues $+1$ and $-1$), traceless and satisfy that $\sigma_i^2 = I$. Together with the identity matrix (which is sometime denoted by $\sigma_0$), they form an orthonormal basis known as the Pauli basis (with respect to the Hilbert-Schmidt product defined as $\braket{A,B} = \trace{(A^{\dagger}B)}$) for any $2 \times 2$ Hermitian matrix. Qubit states can then be expressed in this basis via the density matrix:  
\begin{align}
    \rho = \frac{1}{2}\left(I + \mathbf{r}\cdot\bm{\sigma}\right),
\end{align}
where the vector $\mathbf{r}=(r_x, r_y, r_z)$ is a unit vector called the Bloch vector, and the vector $\bm{\sigma}=(\sigma_x, \sigma_y, \sigma_z)$. The dot product of these two vectors is just a shorthand notation for the expression $\mathbf{r}\cdot\bm{\sigma}= r_x\sigma_x + r_y\sigma_y+ r_z \sigma_z$. Any time-dependent Hamiltonian of a qubit can be expressed as 
\begin{align}
    H(t) = \sum_{i\in\{x,y,z\}} {\alpha_i(t) \sigma_i},
\end{align}
with the time-dependence absorbed in the coefficients $\alpha_i(t)$. 

\subsubsection{Pauli measurements}
As described below, the measurements simulated in the QDataSet are what are known as Pauli measurements. These are formed by taking the expectation value of each Pauli matrix e.g. $\trace{(\rho \sigma_i)}$ for $i \in \{x,y,z\}$ (the identity is omitted). The resultant measurement distributions will typically form labelled data in a machine learning context. As discussed below, measurement distributions are ultimately how various properties of the quantum system are inferred (i.e. via reconstructive inference), such as the characteristics of quantum circuits, evolutionary paths and tomographical quantum state description. As we describe below, measurements in the QDataSet comprise measurements on each eigenstate (six in total) of each Pauli operator by all Pauli operators. [Hermitian] operators have a spectral decomposition in terms of eigenvalues and their corresponding projectors
\begin{align}
    P_0 &= \kz \bz = \frac{1}{2}(I + \sigma_z) = \begin{pmatrix} 
    1 & 0 \\
    0 & 0
    \end{pmatrix}\\
    P_1 &= \kz \bz = \frac{1}{2}(I - \sigma_z) = \begin{pmatrix} 
    0 & 0 \\
    0 & 1
    \end{pmatrix}
\end{align}
thus we can write:
\begin{align}
\sigma_z = 1 \times \begin{pmatrix} 
    1 & 0 \\
    0 & 0
    \end{pmatrix}  - 1 \times \begin{pmatrix} 
    0 & 0 \\
    0 & 1
    \end{pmatrix}  
\end{align}
For example, a Pauli meaurement on a qubit in the $-1$ eigenstate with respect to the $\sigma_z$ operator

\begin{align}
    \trace( \rho \sigma_z^\dagger) =   \trace \left( \begin{pmatrix} 
    0 & 0 \\
    0 & 1
    \end{pmatrix}
    \begin{pmatrix} 
    1 & 0 \\
    0 & -1
    \end{pmatrix} \right) = \trace \begin{pmatrix} 
    0 & 0 \\
    0 & -1
    \end{pmatrix} = -1
\end{align}
which is as expected. The probability of observing $\lambda=-1$ in this state we should expect to be unity (given the state is in the eigenstate):
\begin{align}
   Pr(m=-1) = \trace (P_1^2 \rho) = 1
\end{align}

For $n$-qubit systems (such as two-qubit systems in the QDataSet), Pauli measurements are represented by tensor-products of Pauli operators. For example, a $\sigma_z$ measurement on the first qubit and $\sigma_x$ on the second is represented as:
\begin{align}
    \sigma_z \otimes \sigma_x
\end{align}
In programming matrix notation, this becomes represented as a $4 \times 4$ matrix (tensor):
\begin{align}
    \sigma_z \otimes \sigma_X &=\begin{pmatrix} 
    1 & 0 \\
    0 & -1
    \end{pmatrix} \otimes \begin{pmatrix} 
    0 & 1 \\
    1 & 0
    \end{pmatrix} \\
    &=\begin{pmatrix} 
    1 \times \begin{pmatrix} 
    0 & 1 \\
    1 & 0
    \end{pmatrix}  & 0 \\
    0 & -1 \times \begin{pmatrix} 
    0 & 1 \\
    1 & 0
    \end{pmatrix} 
    \end{pmatrix} \\
    &= \begin{pmatrix} 
    0 & 1 & 0 & 0\\
    1 & 0 & 0 & 0\\
    0 & 0 & 0 & -1\\
    0 & 0 & -1 & 0
    \end{pmatrix} 
\end{align}

\begin{figure}
    \centering
    \includegraphics[width=0.5\textwidth]{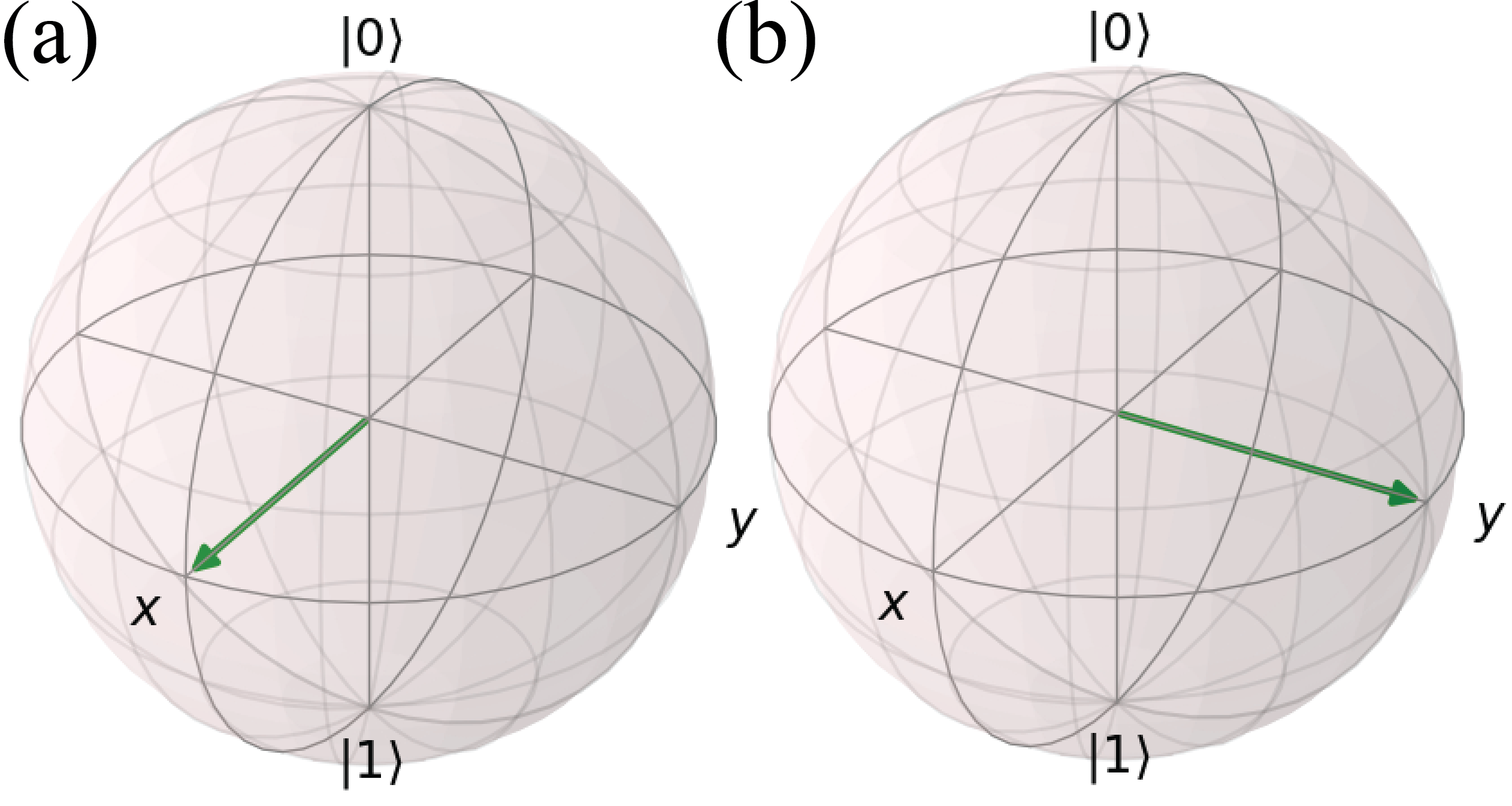}
    \caption{An example of a quantum state rotation on the Bloch sphere. The $\kz,\ko$ indicates the $\sigma_z$-axis, the $X$ and $Y$ the $\sigma_x$ and $\sigma_y$ axes respectively. In (a), the vector is residing in a $+1$ $\sigma_x$ eigenstate. By rotating about the $\sigma_z$ axis by $\pi/4$, the vector is rotated to the right, to the $+1$ $\sigma_y$ eigenstate. A rotation about the $\sigma_Z$ axis by angle $\theta$ is equivalent to the application of the unitary $U(\theta) = \exp(-i \theta_z \sigma_z/2)$.}
    \label{fig:blochrotation}
\end{figure}

The Pauli representation of qubits can be usefully visualised via the Bloch sphere as per Figure \ref{fig:blochrotation}. The axes of the Bloch sphere are the expectation values of the Pauli $\sigma_x, \sigma_y$ and $\sigma_z$ operators respectively. As each Pauli operator has eigenvalues $1$ and $-1$, the eigenvalues can be plotted along axes of the 2-sphere. For a pure (non-decohered) quantum state $\rho$, $|\rho|=\sqrt{r_x^2 + r_y^2 + r_z^2}=1$ (as we require $\trace{\rho^2=1}$), thus $\rho$ is represented on the Bloch 2-sphere as a vector originating at the origin and lying on the surface of the Bloch 2-sphere. The evolution of the qubit i.e. a computation according to unitary evolution can then be represented as rotations of $\rho$ across the Bloch sphere. In noisy contexts, decohered $\rho$ are represented whereby $|\rho|<1$ i.e. the norm of $\rho$ shrinks and $\rho$ no longer resides on the surface. 

For machine learning practitioners, it is useful to appreciate the the operation of Pauli operators $\sigma_x, \sigma_y, \sigma_z$ as the generators of rotations about the respective axes of the Bloch sphere (again, this is just a way of representing the evolution of quantum information). Thus the application of $\sigma_z$ to a qubit is akin to rotating the quantum state vector $\rho$ about the $z$-axis. Figure \ref{fig:blochrotation} Conceptually, a qubit is in a $z$-eigenstate if it is lying directly on either the north ($+1$) or south ($-1$) pole. Rotating about the $z$-axis then is akin to rotating the vector on the spot, thus no change in the quantum states (or eigenvalues) for $\sigma_z$ occurs because the system exhibits symmetry under such transformations. This is similarly the case for $\sigma_x, \sigma_y$ generators with respect to their eigenvalues and eigenvectors. However, rotations by $\sigma_\alpha$ will affect the eigenvalues/vectors in the $\sigma_\beta$ basis where $\alpha \neq \beta$ e.g. a $\sigma_x$ rotation will affect the component of the qubit lying along the $\sigma_z$ axis. Similarly, a $\sigma_z$ rotation of a qubit in a $\sigma_x$ eigenstate will alter that state (shown in (a) and (b) of Figure \ref{fig:blochrotation}). An understanding of Pauli operators and conceptualisation of qubit axes is important to the understanding of the simulated QDataSet. An understanding of symmetries of relevance to qubit evolution (and quantum algorithms) is also beneficial. As we describe below, controls or noise are structured to be applied along particular axes of a qubit and thus can be thought of as a way to control or distortions upon the axial rotations of a qubit effected by the corresponding Pauli generator.

There exist higher dimensional generalization to the Pauli matrices that allow forming orthonormal basis to represent operators in these dimensions. In particular if we have a system of $N$ qubits, then one simple generalization is to form the set $\{\sigma_{i_1}^{(1)} \otimes \sigma_{i_2}^{(2)} \otimes \cdots \sigma_{i_N}^{(N)} \}_{i_j \in \{0,x,y,z\}}$. In other words we take tensor products of the Pauli's which gives a set of size $2^N$. For example, for a two-qubit system we can form the $16-$ element set $\{\sigma_0 \otimes \sigma_0, \sigma_0 \otimes \sigma_x, \sigma_0 \otimes \sigma_y, \sigma_0 \otimes \sigma_z, \sigma_x \otimes \sigma_0, \sigma_x \otimes \sigma_x, \sigma_x \otimes \sigma_y, \sigma_x \otimes \sigma_z, \sigma_y \otimes \sigma_0, \sigma_y \otimes \sigma_x, \sigma_y \otimes \sigma_y, \sigma_y \otimes \sigma_z, \sigma_z \otimes \sigma_0, \sigma_z \otimes \sigma_x, \sigma_z \otimes \sigma_y, \sigma_z \otimes \sigma_z\}$. Moreover, for many use cases, we are interested in the \textit{minimal} number of operators, such as Pauli operators, required to achieve a requisite level of control, such as universal quantum computation. For some machine learning applications, we are interested in minimising the number of controls that must be applied to a quantum system (thus minimising the resources required to control the system). In such cases, we may seek a minimal control algebra or gate set. The minimal number of Pauli operators required to achieve a complete control set of generators for synthesising an arbitrary unitary acting on an $n$ dimensional Hilbert space is given by a bracket-generating set $\Delta \subseteq \mathfrak{su}(2^n)$  \cite{montgomery_tour_2002, perrier_quantum_2020} which can be understood in more complex treatments in the context of Lie algebras and representation theory. Here $\mathfrak{su}(2^n)$ represents the Lie algebra corresponding to the Pauli group SU$(2^n)$, the complete set of generators required to span the $n$ dimensional Hilbert space in the Pauli basis. Given $\Delta$, we can reconstruct the full Pauli basis via the operation of the Lie bracket \cite{swaddle_generating_2017-1}. The QDataSet generators for one- and two-qubit systems are simply one and two (tensor-product) sets of Paulis respectively (i.e. not $\Delta$). For higher-dimensional problems, whether to restrict generators to those within $\Delta$ becomes a consideration for machine learning architectures (see literature on time-optimal quantum control). These generators will typically be used as the tensors or matrices to which classical controls are applied within machine learning architectures.  

For the single qubit system, initial states are the two eigenstates of every Pauli operators. As noted above, the quantum state can be decomposed in the Pauli basis as $\rho_j =\frac{1}{2}(I\pm \sigma_j)$, for $j=1,2,3$. This gives a total of 6 states. We perform the three Pauli measurements on each of these states, resulting in a total of 18 possible combinations. These 18 measurements are important to characterize a qubit system. For two-qubits, it will be similar but now we initialize every individual qubit into the 6 possible eigenstates, and we measure all 15 Pauli observables (we exclude identity). This gives a total of 540 possible combinations. 

\subsubsection{Monte Carlo measurements}

Measurements of the one- and two-qubit systems for the QDataSet are undertaken using Monte Carlo techniques. This means that a random Pauli measurement is undertaken multiple times, with the measurement results averaged in order to provide the resultant measurement distribution for each of the POVM operators. The measurement of the quantum systems is contingent on the noise realisations for each system. For the noiseless case, the Pauli measurements are simply the Monte Carlo averages (expectations) of the Pauli operators. Systems with noise will have one or more noise realisations (applications of noise) applied to them. To account for this, we include two separate sets of measurement distribution. The first the expectation value of the three Pauli operators over all possible initial states for each different noise realisation. Thus for each type of noise, there will be a set of measurement statistics. The second is a set of measurement statistics where we average over all noise realisations for the dataset. Doing so enables algorithms trained using the QDataSet to be more fine-grained in their treatment of noise: in some contexts, while noise profiles may be uncertain, it is clear that the noise is of a certain type, so the first set of measurement statistics may be more applicable. These statistics are given by the the set $\{ V_O \}$ in the QDataSet. For other cases, there is almost no information about noise profiles or their sources, in which case the average over all noise realisations may be more appropriate. This second set of measurements is given by the set $\{E_O \}$.

\subsubsection{Monte Carlo Simulator}
For the benefit of researchers using the QDataSet, we briefly set out a bit more detail of how the datasets were generated. The simulator comprises three main components. The first approximates time-ordered unitary evolution. The second component generates realisations of noise given random parametrisations of the power spectral density (PSD) of the noise. The third component simulates quantum measurement. The simulations are based upon Monte Carlo methods whereby $K$ randomised pulse sequences give rise to noise realisations. The quantum systems are then measured to determine the extent to which the noise realisations affect the expectation values (Pauli measurements). Trial and error indicated a stabilisation of measurement statistics at around $K=500$, thus $K \geq1000$ was chosen for the final simulation run to generate the QDataSet. The final Pauli measurements are then averages over such noise realisations. The parameter $K$ is included for each dataset and example (as described below). For more detail, including useful pseudocode that sets out the relationship between noise realisations, $\beta(t)$ and measurement, see the Supplementary Material in \cite{youssry_characterization_2020}. 

\section{Key parameters}
Table (\ref{tab:params}) sets out an example of parameters for various of the datasets.

\begin{table}
    % \centering
    \begin{tabular}{|c|c|}
    \hline
        Parameter & Value \\
        \hline
        $T$ & 1 \\
        \hline
        $M$ & 1024 \\
        \hline 
        $K$ & 2000 \\
        \hline
        $\Omega$ & 12 \\
        \hline
        $\Omega_1$ & 12 \\
        \hline
        $\Omega_2$ & 10 \\
        \hline
        $n$ & 5 \\
        \hline
        $A_{\text{min}}$ & -100 \\
        \hline
        $A_{\text{max}}$ & 100 \\
        \hline
        $\sigma$ & T/(12M) \\
        \hline
    \end{tabular}
    \caption{Dataset Parameters:  $T$: total time, set to unity for standardisation; $M$: the number of time-steps (discretisations); $K$: the number of noise realisations; $\Omega$: the energy gap for the single qubit case (where subscripts 1 and 2 represent the energy gap for each qubit in the single qubit case); $n$: number of control pulses; $A_{\text{max}},A_{\text{min}}$: maximum and minimum amplitude; $\sigma$: standard deviation of pulse spacing (for Gaussian pulses).}
    \label{tab:params}
\end{table}

\newpage
\section{QDataSet Characteristics}

%======QDataSet filename table

\begin{center}
\begin{longtable}{|p{4cm}||p{13cm}|}
\hline
\multicolumn{2}{|c|}{QDataSet File Descriptions} \\
 \hhline{|==|}
 \textbf{Dataset}  & \textbf{Description}   \\
 \hline
 %=====Gaussian pulse
 G\_1q\_X  & (i) Qubits: one; (ii) Control: $x$-axis, Gaussian; (iii) Noise: none; (iv) No distortion. 
 \\
 \hline
 G\_1q\_X\_D  & (i) Qubits: one; (ii) Control: $x$-axis, Gaussian; (iii) Noise: none; (iv) Distortion. 
 \\
 \hline
 
 G\_1q\_XY  & (i) Qubits: one; (ii) Control: $x$-axis and $y$-axis, Gaussian; (iii) Noise: none; (iv) No distortion. 
 \\
 \hline
 G\_1q\_XY\_D  & (i) Qubits: one; (ii) Control: $x$-axis and $y$-axis, Gaussian; (iii) Noise: none; (iv) Distortion. 
 \\
 \hline
 
  G\_1q\_XY\_XZ\_N1N5  & (i) Qubits: one; (ii) Control: $x$-axis and $y$-axis, Gaussian; (iii) Noise: N1 on $x$-axis, N5 on $z$-axis; (iv) No distortion. 
 \\
 \hline
 G\_1q\_XY\_XZ\_N1N5\_D  & (i) Qubits: one; (ii) Control: $x$-axis and $y$-axis, Gaussian; (iii) Noise: N1 on $x$-axis, N5 on $z$-axis; (iv) No distortion. 
 \\
 \hline
 
 G\_1q\_XY\_XZ\_N1N6  & (i) Qubits: one; (ii) Control: $x$-axis and $y$-axis, Gaussian; (iii) Noise: N1 on $x$-axis, N6 on $z$-axis; (iv) Distortion. 
 \\
 \hline
 G\_1q\_XY\_XZ\_N1N6\_D  & (i) Qubits: one; (ii) Control: $x$-axis and $y$-axis, Gaussian; (iii) Noise: N1 on $x$-axis, N6 on $z$-axis; (iv) No distortion. 
 \\
 \hline
 
  G\_1q\_XY\_XZ\_N3N6  & (i) Qubits: one; (ii) Control: $x$-axis and $y$-axis, Gaussian; (iii) Noise: N1 on $x$-axis, N6 on $z$-axis; (iv) Distortion. 
 \\
 \hline
 G\_1q\_XY\_XZ\_N3N6\_D  & (i) Qubits: one; (ii) Control: $x$-axis and $y$-axis, Gaussian; (iii) Noise: N1 on $x$-axis, N6 on $z$-axis; (iv) No distortion. 
 \\
 \hline
 G\_1q\_X\_Z\_N1  & (i) Qubits: one; (ii) Control: $x$-axis, Gaussian; (iii) Noise: N1 on $z$-axis; (iv) No distortion. 
 \\
 \hline
 G\_1q\_X\_Z\_N1\_D  & (i) Qubits: one; (ii) Control: $x$-axis, Gaussian; (iii) Noise: N1 on $z$-axis; (iv) Distortion. 
 \\
 \hline
 G\_1q\_X\_Z\_N2  & (i) Qubits: one; (ii) Control: $x$-axis, Gaussian; (iii) Noise: N2 on $z$-axis; (iv) No distortion. 
 \\
 \hline
 G\_1q\_X\_Z\_N2\_D  & (i) Qubits: one; (ii) Control: $x$-axis, Gaussian; (iii) Noise: N2 on $z$-axis; (iv) Distortion. 
 \\
 \hline
 G\_1q\_X\_Z\_N3  & (i) Qubits: one; (ii) Control: $x$-axis, Gaussian; (iii) Noise: N3 on $z$-axis; (iv) No distortion. 
 \\
 \hline
 G\_1q\_X\_Z\_N3\_D  & (i) Qubits: one; (ii) Control: $x$-axis, Gaussian; (iii) Noise: N3 on $z$-axis; (iv) Distortion. 
 \\
 \hline
 G\_1q\_X\_Z\_N4  & (i) Qubits: one; (ii) Control: $x$-axis, Gaussian; (iii) Noise: N4 on $z$-axis; (iv) No distortion. 
 \\
 \hline
 G\_1q\_X\_Z\_N4\_D  & (i) Qubits: one; (ii) Control: $x$-axis, Gaussian; (iii) Noise: N4 on $z$-axis; (iv) Distortion. 
 \\
 \hline
 G\_2q\_IX-XI\_IZ-ZI\_N1-N6  & (i) Qubits: two; (ii) Control: $x$-axis on both qubits, Gaussian; (iii)
      Noise: N1 and N6 $z$-axis on each qubit; (iv) No distortion. 
 \\
 \hline
 G\_2q\_IX-XI\_IZ-ZI\_N1-N6\_D  & (i) Qubits: two; (ii) Control: $x$-axis on both qubits, Gaussian; (iii) Noise: N1 and N6 $z$-axis on each qubit; (iv) Distortion. 
 \\
 \hline
 G\_2q\_IX-XI-XX  & (i) Qubits: two; (ii) Control: single $x$-axis control on both qubits and $x$-axis interacting control, Gaussian; (iii) Noise: none; (iv) No distortion. 
 \\
 \hline
 G\_2q\_IX-XI-XX\_D  & (i) Qubits: two; (ii) Control: single $x$-axis control on both qubits and $x$-axis interacting control, Gaussian; (iii) Noise: none; (iv) Distortion. 
 \\
 \hline
 G\_2q\_IX-XI-XX\_IZ-ZI\_N1-N5  & (i) Qubits: two; (ii) Control: single $x$-axis control on both qubits and $x$-axis interacting control, Gaussian; (iii) Noise: N1 and N5 on $z$-axis noise on each qubit; (iv) No distortion. 
 \\
 \hline
 G\_2q\_IX-XI-XX\_IZ-ZI\_N1-N5  & (i) Qubits: two; (ii) Control: single $x$-axis control on both qubits and $x$-axis interacting control, Gaussian; (iii) Noise: N1 and N5 on $z$-axis noise on each qubit; (iv) Distortion. 
 \\
 \hline
 %=========square pulse
 
 S\_1q\_X  & (i) Qubits: one; (ii) Control: $x$-axis, square; (iii) Noise: none; (iv) No distortion. 
 \\
 \hline
 S\_1q\_X\_D  & (i) Qubits: one; (ii) Control: $x$-axis, Gaussquaresian; (iii) Noise: none; (iv) Distortion. 
 \\
 \hline
 
 S\_1q\_XY  & (i) Qubits: one; (ii) Control: $x$-axis and $y$-axis, square; (iii) Noise: none; (iv) No distortion. 
 \\
 \hline
 S\_1q\_XY\_D  & (i) Qubits: one; (ii) Control: $x$-axis and $y$-axis, Gaussquaresian; (iii) Noise: none; (iv) Distortion. 
 \\
 \hline
 
  S\_1q\_XY\_XZ\_N1N5  & (i) Qubits: one; (ii) Control: $x$-axis and $y$-axis, square; (iii) Noise: N1 on $x$-axis, N5 on $z$-axis; (iv) No distortion. 
 \\
 \hline
 S\_1q\_XY\_XZ\_N1N5\_D  & (i) Qubits: one; (ii) Control: $x$-axis and $y$-axis, Gaussian; (iii) Noise: N1 on $x$-axis, N5 on $z$-axis; (iv) No distortion. 
 \\
 \hline
 
 S\_1q\_XY\_XZ\_N1N6  & (i) Qubits: one; (ii) Control: $x$-axis and $y$-axis, square; (iii) Noise: N1 on $x$-axis, N6 on $z$-axis; (iv) Distortion. 
 \\
 \hline
 S\_1q\_XY\_XZ\_N1N6\_D  & (i) Qubits: one; (ii) Control: $x$-axis and $y$-axis, square; (iii) Noise: N1 on $x$-axis, N6 on $z$-axis; (iv) No distortion. 
 \\
 \hline
 
  S\_1q\_XY\_XZ\_N3N6  & (i) Qubits: one; (ii) Control: $x$-axis and $y$-axis, square; (iii) Noise: N1 on $x$-axis, N6 on $z$-axis; (iv) Distortion. 
 \\
 \hline
 S\_1q\_XY\_XZ\_N3N6\_D  & (i) Qubits: one; (ii) Control: $x$-axis and $y$-axis, square; (iii) Noise: N1 on $x$-axis, N6 on $z$-axis; (iv) No distortion. 
 \\
 \hline
 S\_1q\_X\_Z\_N1  & (i) Qubits: one; (ii) Control: $x$-axis, square; (iii) Noise: N1 on $z$-axis; (iv) No distortion. 
 \\
 \hline
 S\_1q\_X\_Z\_N1\_D  & (i) Qubits: one; (ii) Control: $x$-axis, square; (iii) Noise: N1 on $z$-axis; (iv) Distortion. 
 \\
 \hline
 S\_1q\_X\_Z\_N2  & (i) Qubits: one; (ii) Control: $x$-axis, square; (iii) Noise: N2 on $z$-axis; (iv) No distortion. 
 \\
 \hline
 G\_1q\_X\_Z\_N2\_D  & (i) Qubits: one; (ii) Control: $x$-axis, Gaussian; (iii) Noise: N2 on $z$-axis; (iv) Distortion. 
 \\
 \hline
 S\_1q\_X\_Z\_N3  & (i) Qubits: one; (ii) Control: $x$-axis, square; (iii) Noise: N3 on $z$-axis; (iv) No distortion. 
 \\
 \hline
 S\_1q\_X\_Z\_N3\_D  & (i) Qubits: one; (ii) Control: $x$-axis, square; (iii) Noise: N3 on $z$-axis; (iv) Distortion. 
 \\
 \hline
 S\_1q\_X\_Z\_N4  & (i) Qubits: one; (ii) Control: $x$-axis, square; (iii) Noise: N4 on $z$-axis; (iv) No distortion. 
 \\
 \hline
 S\_1q\_X\_Z\_N4\_D  & (i) Qubits: one; (ii) Control: $x$-axis, square; (iii) Noise: N4 on $z$-axis; (iv) Distortion. 
 \\
 \hline
 S\_2q\_IX-XI\_IZ-ZI\_N1-N6  & (i) Qubits: two; (ii) Control: $x$-axis on both qubits, square; (iii) Noise: N1 and N6 $z$-axis on each qubit; (iv) No distortion. 
 \\
 \hline
 S\_2q\_IX-XI\_IZ-ZI\_N1-N6\_D  & (i) Qubits: two; (ii) Control: $x$-axis on both qubits, square; (iii) Noise: N1 and N6 $z$-axis on each qubit; (iv) Distortion. 
 \\
 \hline
 S\_2q\_IX-XI-XX  & (i) Qubits: two; (ii) Control: single $x$-axis control on both qubits and $x$-axis interacting control, square; (iii) Noise: none; (iv) No distortion. 
 \\
 \hline
 S\_2q\_IX-XI-XX\_D  & (i) Qubits: two; (ii) Control: single $x$-axis control on both qubits and $x$-axis interacting control, square; (iii) Noise: none; (iv) Distortion. 
 \\
 \hline
 
 S\_2q\_IX-XI-XX\_IZ-ZI\_N1-N5  & (i) Qubits: two; (ii) Control: $x$-axis on both qubits and $x$-axis interacting control, square; (iii) Noise: N1 and N5 $z$-axis on each qubit; (iv) No distortion. 
 \\
 \hline
 S\_2q\_IX-XI-XX\_IZ-ZI\_N1-N5\_D  & (i) Qubits: two; (ii) Control: $x$-axis on both qubits and $x$-axis interacting control, square; (iii) Noise: N1 and N5 $z$-axis on each qubit; (iv) Distortion. 
 \\
 \hline
 
 S\_2q\_IX-XI-XX\_IZ-ZI\_N1-N6  & (i) Qubits: two; (ii) Control: $x$-axis on both qubits and $x$-axis interacting control, square; (iii) Noise: N1 and N6 $z$-axis on each qubit; (iv) No distortion. 
 \\
 \hline
 S\_2q\_IX-XI-XX\_IZ-ZI\_N1-N6\_D  & (i) Qubits: two; (ii) Control: $x$-axis on both qubits and $x$-axis interacting control, square; (iii) Noise: N1 and N6 $z$-axis on each qubit; (iv) Distortion. 
 \\
 \hline
 
 \caption{QDataSet File Description. The left column identifies each dataset in the respective QDataSet examples while the description column describes the profile of the dataset in terms of (i) number of qubits, (ii) axis of control and pulse wave-form (iii) axis and type of noise and (iv) whether distortion is present or absent. }
\label{table:datasetproperties}
\end{longtable}
\end{center}

\newpage

\end{document}